\documentclass[longauth]{aa} 
\usepackage{graphicx}
\usepackage{txfonts}

\newcommand{\kms}{km\,s$^{-1}$}
\newcommand{\hei}[0]{\ion{He}{i} D$_{3}$}
\newcommand{\si}[0]{\ion{Na}{i} D$_{1}$}
\newcommand{\sii}[0]{\ion{Na}{i} D$_{2}$}
\newcommand{\ha}[0]{H${\alpha}$}
\newcommand{\ca}[0]{\ion{Ca}{ii} IRT}
\newcommand{\prot}[0]{\emph{P}$_{\mathrm{rot}}$}

\begin{document} 

\title{Simultaneous photometric and CARMENES spectroscopic monitoring of fast-rotating M dwarf GJ~3270}

\subtitle{Discovery of a post-flare corotating feature}

\titlerunning{}
\authorrunning{Johnson et al.}
\author{E.\,N.~Johnson\inst{1,2}
  \and S.~Czesla\inst{3} 
  \and B.~Fuhrmeister\inst{3}
  \and P.~Sch\"ofer\inst{1}
  \and Y. Shan\inst{1}
  \and C.~Cardona Guillén\inst{4,5}
  \and A.~Reiners\inst{1}
  \and S.~V.~Jeffers\inst{2} 
  \and S.~Lalitha\inst{6}
  \and R.~Luque\inst{4,5}
  \and E.~Rodr\'iguez\inst{7}
  \and V.\,J.\,S.~B\'ejar\inst{4,5}
  \and J.\,A.~Caballero\inst{8}
  \and L.~Tal-Or\inst{9,1}
  \and M.~Zechmeister\inst{1}
  \and I.~Ribas\inst{10,11}
  \and P.\,J.~Amado\inst{7}
  \and A.~Quirrenbach\inst{12}
  \and M.~Cort\'es-Contreras\inst{8}
  \and S.~Dreizler\inst{1}
  \and A.~Fukui\inst{13,4}
  \and M.\,J.~López-González\inst{7}
  \and A.\,P.~Hatzes\inst{14}
  \and Th.~Henning\inst{15}
  \and A. Kaminski\inst{12}
  \and M.~K\"urster\inst{15}
  \and M.~Lafarga\inst{10,11}
  \and D.~Montes\inst{16}
  \and J.\,C.~Morales\inst{10,11}
  \and F.~Murgas\inst{4,5}
  \and N.~Narita\inst{17,18,19,4}
  \and E.~Pall\'e\inst{4,5}
  \and H.~Parviainen\inst{4,5}
  \and S.~Pedraz\inst{20}
  \and D.~Pollacco\inst{21}
  \and A.~Sota\inst{7}
  }

\institute{Institut f\"ur Astrophysik, Friedrich-Hund-Platz 1, D-37077 G\"ottingen, Germany\\
  \email{erik.johnson@uni.goettingen.de}
          \and 
        Max Planck Institute for Solar System Research, Justus-von-Liebig-Weg 3, 37077 G\"ottingen, Germany.
        \and
        Hamburger Sternwarte, Universit\"at Hamburg, Gojenbergsweg 112, D-21029 Hamburg, Germany
        \and 
        Instituto de Astrof\'{\i}sica de Canarias, c/ V\'{\i}a L\'actea s/n, E-38205 La Laguna, Tenerife, Spain 
        \and
        Departamento de Astrof\'{\i}sica, Universidad de La Laguna, E-38206 Tenerife, Spain 
        \and
        School of Physics \& Astronomy, University of Birmingham, Edgbaston, Birmingham B15 2TT, UK 
        \and
        Instituto de Astrof\'isica de Andaluc\'ia (CSIC), Glorieta de la Astronom\'ia s/n, E-18008 Granada, Spain 
        \and 
        Centro de Astrobiolog\'{\i}a (CSIC-INTA), ESAC, Camino Bajo del Castillo s/n, E-28692 Villanueva de la Ca\~nada, Madrid, Spain 
        \and 
        Department of Physics, Ariel University, Ariel 40700, Israel 
        \and
        Institut de Ci\`encies de l'Espai (ICE, CSIC), Campus UAB, c/ de Can Magrans s/n, E-08193 Bellaterra, Barcelona, Spain 
        \and
        Institut d'Estudis Espacials de Catalunya (IEEC), E-08034 Barcelona, Spain
        \and
        Landessternwarte, Zentrum f\"ur Astronomie der Universit\"at Heidelberg, K\"onigstuhl 12, D-69117 Heidelberg, Germany 
        \and
        Department of Earth and Planetary Science, Graduate School of Science, The University of Tokyo, 7-3-1 Hongo, Bunkyo-ku, Tokyo 113-0033, Japan 
        \and
        Th\"uringer Landessternwarte Tautenburg, Sternwarte 5, D-07778 Tautenburg, Germany 
        \and
        Max-Planck-Institut f\"ur Astronomie, K\"onigstuhl 17, D-69117 Heidelberg, Germany 
        \and
        Departamento de F{\'i}sica de la Tierra y Astrof{\'i}sica \& IPARCOS-UCM (Instituto de F\'{i}sica de Part\'{i}culas y del Cosmos de la UCM),
        Facultad de Ciencias F{\'i}sicas, Universidad Complutense de Madrid, E-28040 Madrid, Spain 
        \and
        Komaba Institute for Science, The University of Tokyo, 3-8-1 Komaba, Meguro, Tokyo 153-8902, Japan 
        \and
        JST, PRESTO, 3-8-1 Komaba, Meguro, Tokyo 153-8902, Japan 
        \and
        Astrobiology Center, 2-21-1 Osawa, Mitaka, Tokyo 181-8588, Japan 
        \and
        Centro Astron\'omico Hispano-Alem\'an (MPG-CSIC), Observatorio Astron\'omico de Calar Alto, Sierra de los Filabres, E-04550 G\'ergal, Almer\'{\i}a, Spain 
        \and
        Department of Physics, University of Warwick, Gibbet Hill Road, Coventry CV4 7AL, UK 
       }

   \date{}
   
\date{Received 17 December 2020 / Accepted 23 March 2021}

 \abstract{Active M dwarfs frequently exhibit large flares, which can pose an existential threat to the habitability of any planet in orbit in addition to making said planets more difficult to detect. M dwarfs do not lose angular momentum as easily as earlier-type stars, which maintain the high levels of stellar activity for far longer. Studying young, fast-rotating M dwarfs is key to understanding their near stellar environment and the evolution of activity.}
 {We study stellar activity on the fast-rotating M~dwarf GJ~3270.}
 {We analyzed dedicated high cadence, simultaneous, photometric and high-resolution spectroscopic observations obtained with CARMENES of GJ~3270 over 7.7 h, covering a total of eight flares of which two are strong enough to facilitate a detailed analysis. We consult the \textit{TESS} data, obtained in the month prior to our own observations, to study rotational modulation and to compare the \textit{TESS} flares to those observed in our campaign.}
 {The \textit{TESS} data exhibit rotational modulation with a period of $0.37$~d. The strongest flare covered by our observing campaign released a total energy of about $3.6\times 10^{32}$\,erg, putting it close to the superflare regime. This flare is visible in the $B$,$V$, $r$, $i$, and $z$ photometric bands, which allows us to determine a peak temperature of about $10\,000$\,K. The flare also leaves clear marks in the spectral time series. In particular,
 we observe an evolving, mainly blue asymmetry in chromospheric lines, which we attribute to a post-flare, corotating feature. To our knowledge this is the first time such a feature has been seen on a star other than our Sun.}
 {Our photometric and spectroscopic time series covers the eruption of a strong flare followed up by a corotating feature analogous to a post-flare arcadal loop on the Sun with a possible failed ejection of material.} 

   
   \maketitle

\section{Introduction}

As a result of their ubiquity, low mass, and close-in habitable zones, M dwarfs have garnered the interest of exoplanet surveys hunting Earth-like analogs. Some of these stars, however, are also known to have exceptional levels of stellar activity \citep{Gizis2000,Khodachenko2007,Yelle2008, O'Malley-James2017,Guarcello2019}. These high levels of stellar activity cannot only make planet detection more difficult, but also call into question the habitability of any planets found around these stars \citep{Johnstone2019,Tilley2019}. The ionizing radiation and high energy particles released can erode or completely strip the atmosphere of an otherwise habitable planet. This process is particularly concerning for planets around M dwarfs because the habitable zone of these stars is much closer in. Particularly energetic events have been proposed as triggers of extinction events on Earth \citep{Lingam2017}. Therefore, knowing the frequency, energy, and history of these events on the host star is critical to understanding the habitability potential of a given exoplanet. 

Stellar activity manifests itself on our Sun most prominently in the form of sunspots, plages, flares, and coronal mass ejections \citep[CMEs --][]{Strassmeier1993,benz_2010}. Stellar activity is usually more extreme in younger, faster-rotating stars \citep{Appenzeller1989,Kiraga2007,Newton2016,Guarcello2019}. Additionally the proportion of active to quiet stars in the M spectral type is higher than in other types of stars \citep{West2008,Reiners2012,Jeffers2018}. This effect is even more pronounced for late M dwarfs. It has been proposed, for M dwarfs later than $\sim$M4, that this is due to the geometry of a the magnetic field of a star, which prevents ejection of material and inhibits the magnetic breaking of the star and its transition to a lower activity state \citep{Barnes2003,Reiners2012b}.

While starspots on M dwarfs can often be studied from rotational modulation in photometric time series \citep{Kron1952,Barnes2015}, the most noticeable feature of stellar activity in either photometry or spectroscopy are stellar flares \citep{Budding1977}. Stellar flares result from a release of energy caused by magnetic reconnection in the upper atmosphere \citep{Hawley1991,Haisch1991,Hilton2010,benz_2010}. This reconnection forces free electrons to follow the magnetic field lines into the chromosphere and photosphere. In the chromosphere, the release of X-rays and enhancement in the chromospheric lines is commonly observed. Upward flows of chromospheric material can also occur as heated material rises into the upper atmosphere. This phenomenon is referred to as chromospheric evaporation \citep{Fisher1985,Abbett1999}.  The photosphere reacts by extremely rapid increase in brightness in the affected area (impulsive phase) followed by an exponential decay back to pre-flare brightness (decay phase) once the electron bombardment has ceased \citep{Segura2009}. The decay phase may last minutes to hours and in very rare cases days \citep{Osten2016,Kuerster1996}. Post-flare arcades and additional minor reconnection events are common during this phase \citep{Gopalswamy2015}. In cool stars flares are more noticeable at shorter wavelengths owing to the contrast of the typical temperatures of flares of $\sim$$10^4$\,K \citep{Kowalski2018,Fuhrmeister2018} and the host star of $\sim$$10^3$\,K. As the flare-affected region cools during the decay phase, this contrast fades, thereby leading to a change in the continuum slope over the course of the flare duration \citep{Segura2009}. 

In spectra, flares are usually detected through enhancement of chromospheric lines, particularly the Balmer lines and those of singly ionized calcium \citep{Hawley1991,Crespo2006,Fuhrmeister2008,Schmidt2011,Fuhrmeister2018}. As opposed to the photometric flare signature of a near-immediate peak at the flare onset, spectroscopically observed flares may not have a peak for many tens of minutes into the event \citep{benz_2010}. Line profiles of chromospheric lines can also undergo broadening and exhibit both red and blue asymmetries in response to a flare \citep{Fuhrmeister2018}.

Line asymmetries are thought to vary during the course of a flare. However as a consequence of the random nature of observing a  stellar flare, the most common detection is through chance observations during a survey. By their very nature these observations only show a moment in time of the progression of the flare. It is therefore difficult to ascertain in which phase an observation catches the flare, making the assignment of a phase to any observed line asymmetry impossible. In general blue asymmetries are assumed to occur in the pre-flare or rise phase and are indicative of chromospheric evaporation or other bulk upward plasma motions. Red asymmetries, on the other hand, are thought to be associated with coronal rain and the decay phase \citep{Fuhrmeister2018}. 

The energies of stellar flares can vary dramatically with the magnitude of the flare and wavelength. The most energetic flares can emit 10$^{37}$\,erg in X-rays that can be an order of magnitude more energetic than that observed in visible wavelengths for the same flare \citep{Kuerster1996}. \citet{Gunther2020} estimated the bolometric energy of the largest M dwarf flares to be 10$^{36.9}$\,erg. These estimates, however, are usually based on the assumption that the flare is a blackbody, which may not be a good approximation.  On the Sun the largest flares are three orders of magnitude lower in X-rays \citep{Kane2005}. The total energy released in the Carrington Event, the most powerful flare yet recorded, was estimated to be \textasciitilde10$^{33}$\,erg \citep{Aulanier2013}.  The smallest solar flares have been reported with energies as low as 10$^{23}$\,erg \citep{Parnell2000}.

The largest, longest-lasting solar flares are frequently associated with a CME. The velocity of this ejected mass can vary from 60 to 3200\,\kms\ with masses on the order of $10^{12}$\,kg  \citep{benz_2010}. While CMEs are relatively easy to detect on our Sun, particularly if they directly impact Earth, they are far more difficult to detect on other stars and none have yet been conclusively identified \citep{Vida2019, Leitzinger2020}. This primarily results from their diffuse nature and being outshined by the host star. Therefore, CMEs are easiest to observe in shorter wavelengths where the contrast is the highest. Coronal mass ejections are thought to produce large, asymmetric blue line asymmetries in Balmer lines as detectable indicators \citep{Vida2019}. If the shift in the asymmetry corresponds to a velocity of at least 10\,\% of the stellar escape velocity, we can be reasonably confident that a CME has occurred. These CMEs are frequently associated with prominence ejections. \citet{Munro1979} found that as much as 70\,\% of solar CMEs have an ejected prominence at their core. It is known that the mass of a prominence depends on the strength of the magnetic field of the host star \citep{DAngelo2018}. M dwarfs are known to have much stronger magnetic fields than the Sun \citep{Shulyak2019} and thereby can presumably host much larger prominences. \citet{Cho2016} detect a large prominence prior to a flaring event on the Sun using high cadence spectroscopy.

While flaring is fundamentally random in nature, the odds of observing a flare increases when observing the more active fast-rotating stars due to the rotation-activity relation. The MEarth survey identified a number of stars whose rotational periods are thought to be less than a day \citep{Berta2012}. One of these stars, GJ~3270, is a M4.5\,V star with a $v\sin{i}$ greater than 30\,\kms and a rotation period shorter than 10\,h \citep[e.g.,][]{West2015,Kasseli2018}.

In this paper, we analyze a series of flares that were observed on the ultra-fast-rotating M dwarf GJ~3270, which we observed on 15 December 2018, utilizing high cadence, simultaneous spectroscopy and photometry. 
In Section~\ref{Sec:datred} we provide details on the instruments and the reduction of the data. In Section~\ref{sec:stellarPars} we discuss the stellar parameters of GJ~3270. Section~\ref{sec:Analysis} we introduce the methods used to analyze the data. In Section~\ref{sec:Results} we present the results of our analysis, then discuss these results in Section~\ref{sec:discussion}.

\section{Observations and data reduction}
\label{Sec:datred}

We present the instruments and data reduction used in this paper. The simultaneous, ground-based, photometry is discussed first followed by the long-baseline SuperWASP data. We then discuss the \textit{TESS} data reduction followed up by the spectroscopic data provided by CARMENES.

\subsection{Ground-based photometry}
\label{sec:GBP}

We obtained multiband photometry of GJ~3270
simultaneously with the MuSCAT2 instrument,
mounted at the 1.52\,m Telescopio Carlos S\'achez in the Teide Observatory \citep{muscat2},
and the T150 and T90 Ritchie-Chr\'etien telescopes of the
Observatorio de Sierra Nevada (SNO).
The MuSCAT2 instrument has a field of view (FOV) of
$7.4\times 7.4$\,arcmin. This instrument was designed to carry
out multicolor simultaneous photometry. In our run, we used the
$r$ (full width at half maximum; FWHM: 1240\,\AA, henceforth $r$), $i$ (FWHM: 1303\,\AA, henceforth $i$), and $z_s$ (FWHM: 2558\,\AA) bands, which we refer to as $r$, $i$, and $z$ bands
in the following. The data that we utilized were preprocessed using the MuSCAT2 data pipeline, detailed in \citet{Parviainen2020}. 

The T150 and T90 telescopes at SNO were used to obtain 
simultaneous photometry in the
Johnson $B$ (FWHM: 781\,\AA) and $V$ (FWHM: 991\,\AA) filters. The telescopes
are equipped with 
similar CCD cameras (VersArray 2k\,$\times$\,2k). Their FOVs are 7.9\,$\times$\,7.9\,arcmin$^2$ and 
13.2\,$\times$\,13.2\,arcmin$^2$, respectively \citep{rodriguez_2010}.
During the readout, we
applied 2\,$\times$\,2 binning for the T150 camera and no binning
for the T90 camera.

\begin{figure}
    \centering
    \includegraphics[width=0.5\textwidth, angle=0]{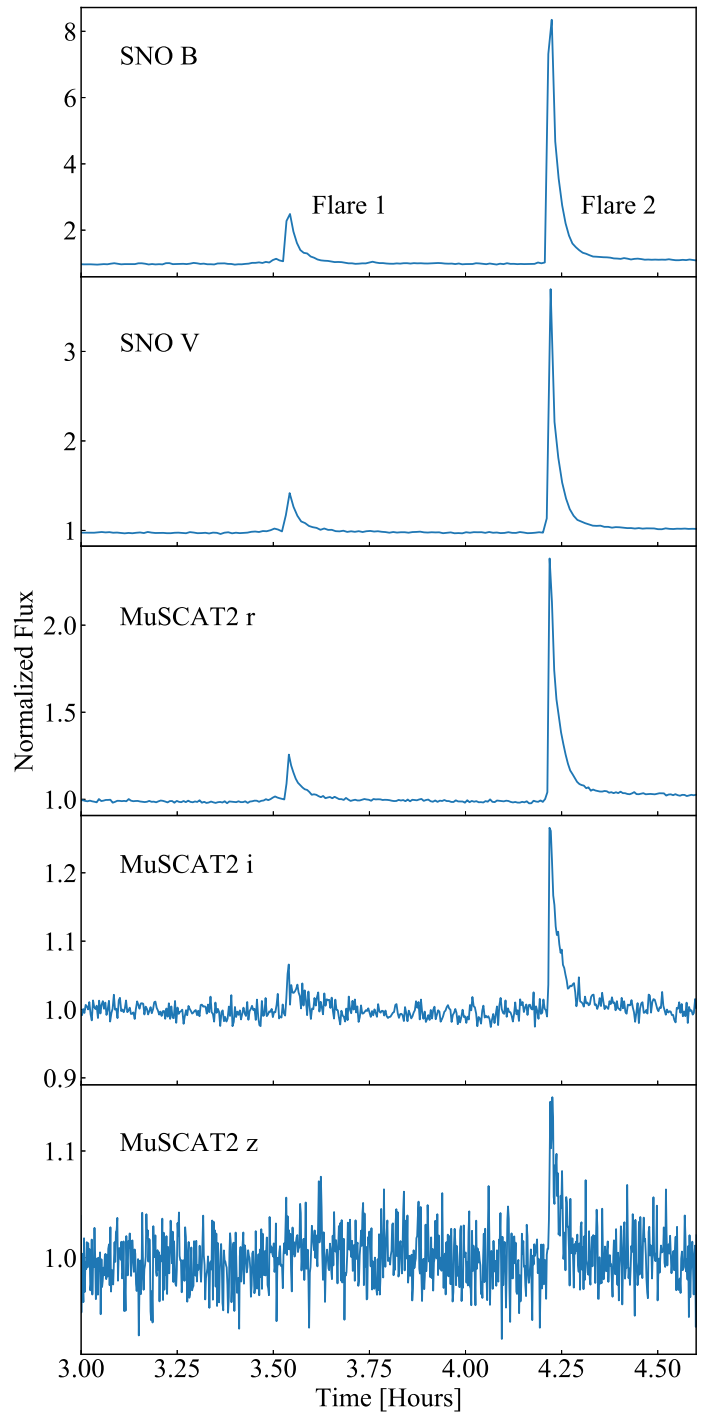}
    \caption{The SNO observations normalized light curves in $V$ and $B$ band are shown in the top two panels. The normalized light curves of the MuSCAT2 data, in $r$, $i$, and $z$ bands, are represented in the bottom three panels.  }
    \label{fig:muscatsno}
\end{figure}

Each CCD frame was corrected for bias and flat field and, 
subsequently, light curves were extracted by
applying synthetic aperture photometry.
All frames cover 
a number of suitable comparison stars for differential photometry.
Different aperture sizes were tested to choose the best size for our observations. The normalization was done by dividing the light curve by its median value. The start time, duration, and exposure times of each photometric run are given in Table~\ref{TB:Ptable}.
Excerpts of the final normalized light curves, showing the two most prominent flaring events,
are shown in Fig~\ref{fig:muscatsno}.

\begin{table} 
    \caption{Start time, duration, and exposure time
     at time of flare of photometric observations. SNO: S, MuSCAT2: M.}
    \label{TB:Ptable} 
    \centering 
    \begin{tabular}{lcccc} 
        \hline \hline
        \noalign{\smallskip}
        Filter & Start & Duration & Exp. time & Total Obs\#\\
               & JD    & [h]      & [s] \\
        \noalign{\smallskip}
        \hline
        \noalign{\smallskip}
        S $B$ & 2458468.304 & 5.0 &  30$^a$ & 379\\
        S $V$ & 2458468.298 & 8.6 &  30$^b$ & 669\\
        M $r$ & 2458468.398 & 5.59 & 22 & 941\\
        M $i$ & 2458468.399 & 5.59 & 12 & 1755\\
        M $z$ & 2458468.399 & 5.59 & 6 & 3093\\
        \noalign{\smallskip}
        \hline 
    \end{tabular} 
\tablefoot{
\tablefoottext{a}{60\,s during the first 3 hs;}
\tablefoottext{b}{100\,s during the first and 60\,s for
the following 2 h.}
}
\end{table}

The Super-Wide Angle Search for Planets (SuperWASP, \citealt{Pollacco06}) survey is a transiting planet survey conducted from two robotic observatories (located in La Palma, Spain, and Sutherland, South Africa), each with a setup of eight wide-angle cameras. The observations are done through a broadband filter covering 400--700\,nm. GJ 3270 was monitored by the SuperWASP program from 2008 to 2014, culminating in $\sim$57\,000 observations over six seasons, each lasting about three months. Data were reduced by the SuperWASP team and detrended using methods designed to preserve variations of astrophysical origin, as detailed in \citet{Tamuz05}. As we utilized SuperWASP data for the sole purpose of analyzing dominant periodicities, associated with the stellar rotation and not for flaring analysis, we filtered the SuperWASP light curves iteratively to remove $4.0\sigma$ outliers.

\subsection{Space-based \textit{TESS} photometry}
\begin{figure*}
     \centering
    \includegraphics[width=0.99\textwidth,angle=0]{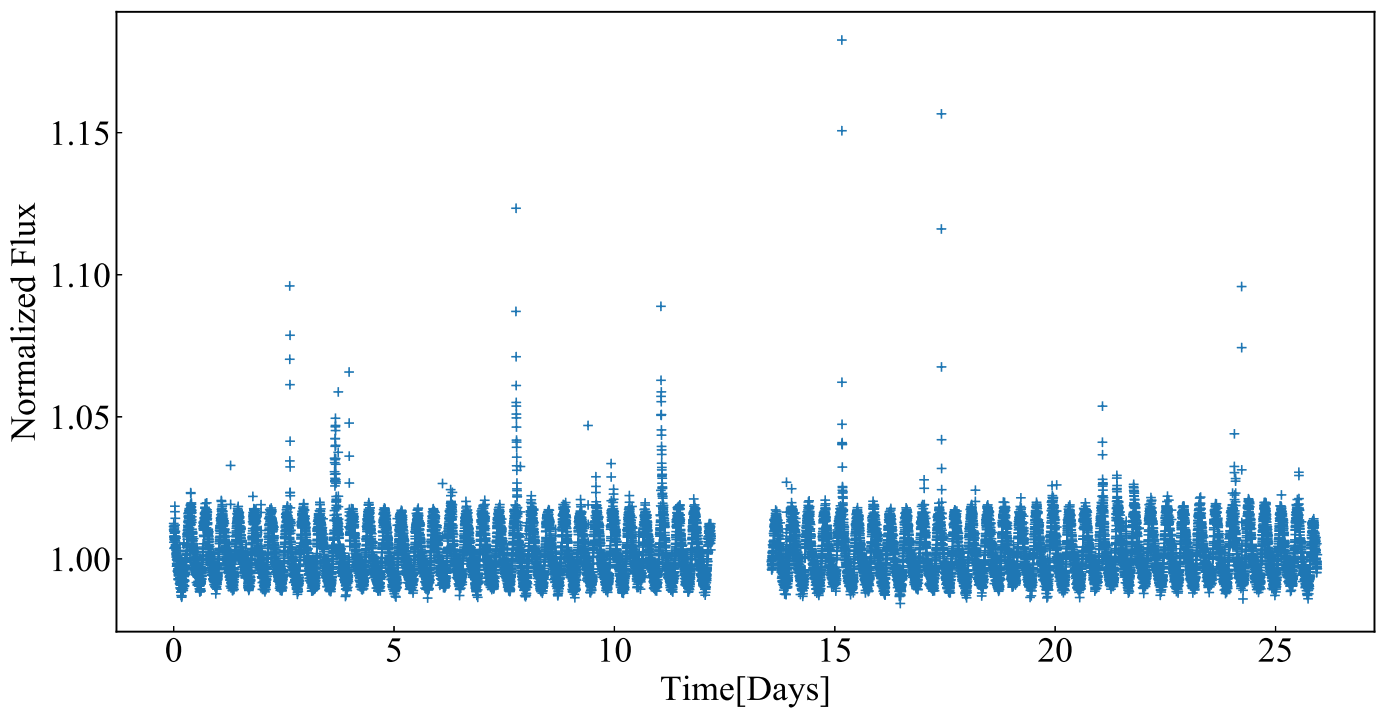}
    \caption{\textit{TESS} light curve of GJ~3270 from sector 5 observations. At least 22 flares occurred during this time span. Rotation of \textasciitilde0.3 d can be seen by inspection of the non-flaring light curve.}
    \label{fig:TESS_ccLC}
\end{figure*}

GJ~3270 was observed in Sector~5 by the {\em Transiting Exoplanet Survey Satellite} (\textit{TESS}; \citealt{TESS_ricker_2015}) in two-minute cadence mode between 15 November and 11 December 2018. These observations ended five days prior to the beginning of our campaign. We used the \textit{TESS} light curves available at Mikulski Archive for Space Telescopes.Utilizing the PDCSAP data, we removed the data points flagged as low-quality by the \textit{TESS} pipeline \citep{Jenkins2016} prior to our analysis. The \textit{TESS} light curve is given in Fig.~\ref{fig:TESS_ccLC}.

\subsection{CARMENES spectra}

CARMENES\footnote{Calar Alto high-Resolution search for M dwarfs with Exoearths with Near-infrared and optical \'{E}chelle Spectrographs.} is a fiber-fed, highly stabilized spectrograph mounted at the Calar Alto 3.5\,m telescope. The instrument has a visual (VIS) and near-infrared (NIR) channel, which are operated simultaneously \citep{Quirrenbach2016}. The VIS channel operates between 520 nm and 960\,nm and the NIR channel between 960 nm and 1710\,nm at spectral resolutions of 94,600 and 80,400, respectively. 

Our observations of GJ~3270 were carried out on 15 December 2018 and comprise 28 VIS and NIR spectra, which each have an exposure time of 15\,min. The spectral time series covers a total of 7.7 h.
All spectra were reduced using the {\tt caracal} pipeline, which
relies on the flat-relative optimal extraction \citep{Zechmeister2014,Caballero2016b}.
In Table ~\ref{TB:timetable}, we give the central time of the observations and the duration since the first observation, and assign an
observation number, which is used to refer to the spectra in the following.

\section{Stellar parameters}
\label{sec:stellarPars}
GJ 3270 has been placed in several young associations. In particular, it was proposed to be a member of the AB Doradus moving group by  \citet{Bell2015}. \citet{Contreras2017} proposed GJ~3270 as a member of the Local Association, also known as the Pleiades moving group \citep{Eggen1983}. These groups range in age from 20 Myr to 300 Myr. The {\em ROSAT} All Sky Survey measured X-ray emission to be $\log(L_x) = 28.3$\,erg\,s$^{-1}$ \citep{Voges1999}. This value is too low for the younger groups but is compatible for those from similar objects in AB Dor. Lithium 6708\,\AA\ was not detected in our spectra. This indicates that the age of GJ~3270 must be greater than about 50\,Myr \citep{Zickgraf2005}. Therefore our age range for GJ~3270 is 50 Myr to 300 Myr with a most probable age of $\sim$150\,Myr as a member of AB Doradus. In color-magnitude diagrams GJ~3270 is not significantly over-luminous \citep{Cifuentes2020}. This indicates that it is nearly, or already on, the main sequence, as is expected for low-mass stars older than 100\,Myr. Therefore, we can use main-sequence relations to determine its stellar parameters.

Using multiwavelength photometry from the blue optical to the mid-infrared, \citet{Cifuentes2020} estimate a $T_{\rm eff}$ of 3100$\pm$50\,K for of GJ~3270. Using magnitude values only from \citet{UCAC4} and the color-temperature relations from \citet{AstrophysicalQuantities}, \citet{NewLightDarkStars}, and \citet{Mamajek2013}, we were able to confirm this value. However, owing the fast rotation and youth of GJ~3270, we assume a conservative $\pm$200\,K uncertainty when using $T_{\rm eff}$ in calculations. We chose a PHOENIX model spectra \citep{Husser2013} with $T_{\rm eff}$ = 3100\,K, $\log{g}=5.0$, and solar metallicity for our template spectra.

\citet{Cifuentes2020} also estimate the luminosity of GJ~3270 to be 0.00642$\pm$0.00003\,$L_{\odot}$. With the mass-luminosity relationship in Eq.~\ref{Eq.:ML} \citep{Schweitzer2019},
\begin{equation}
    \frac{L}{L_{\odot}}=0.163\left (\frac{M}{M_{\odot}}\right)^{2.22\pm0.22} \; ,
    \label{Eq.:ML}
\end{equation}
we determine the mass to be 0.25$\pm$0.07\,$M_{\odot}$, which is consistent with the findings of \citet{Cifuentes2020}.

\begin{figure}
    \centering
    \includegraphics[width=0.5\textwidth, angle=0]{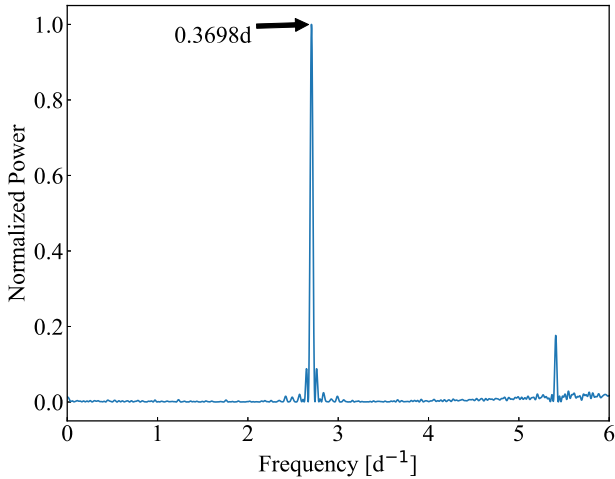}
    \caption{Generalized periodogram of combined TESS and SNO $V$ data. The power level of a 10$^{-3}$ FAP is 0.0219. }
    \label{fig:periodogram}
\end{figure}

\begin{figure*}
     \centering
     \includegraphics[width=0.99\textwidth,angle=0]{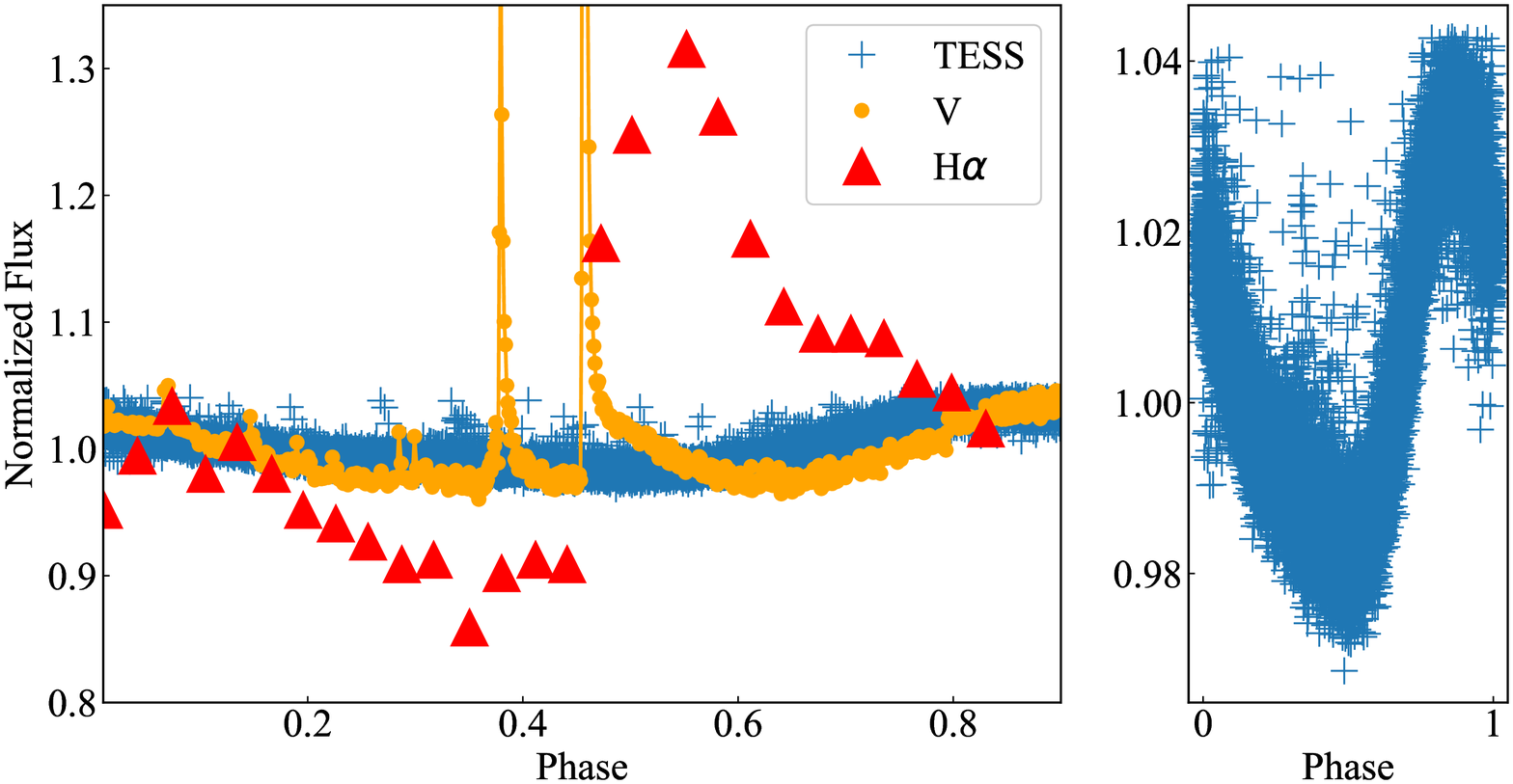}
    \caption{\textit{Left}: Comparison of the phase-folded \textit{TESS} flare-removed light curve (blue crosses) and SNO $V$ (orange circles) photometric data with CARMENES \ha\ $I/I_r$ (red triangles) for flares 1 and 2. The \ha\ values have been normalized by their median value to properly relate them to the \textit{TESS} and SNO $V$ values. 
    \textit{Right}: Phase-folded, flare-removed light curve of \textit{TESS} observations of GJ~3270. }
    \label{fig:TESS_VH}
\end{figure*}

\begin{figure}
     \centering
     \includegraphics[width=0.45\textwidth,angle=0]{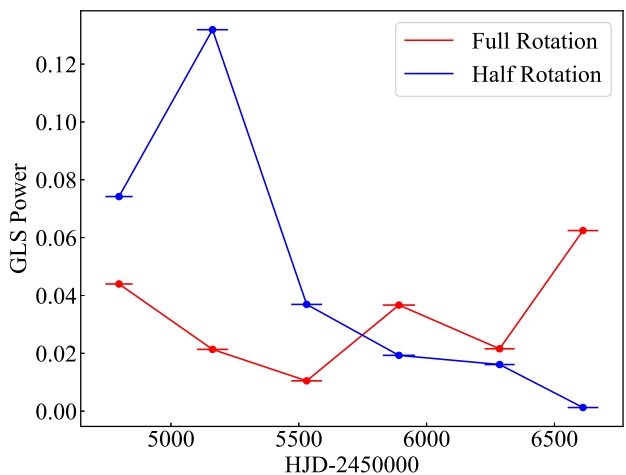}
     \vspace{+0.5cm}
    \caption{Comparison of the GLS power \citep{Zechmeister_Kuerster_2009} of the full- and half-rotation period of GJ~3270 from SuperWASP data. The error bars denote observing seasons from which the GLS periodograms were generated.  }
    \label{fig:SWASP_GLS}
\end{figure}

Five days prior to our SNO and MuSCAT2 observations, \textit{TESS} ended its observation of GJ~3270.
We searched for periodic signals in the combined data set of the \textit{TESS}, SNO, and MuSCAT2 photometric data using the generalized Lomb–Scargle periodogram  \citep[GLS,][]{Zechmeister_Kuerster_2009}. 
The resulting power spectrum is shown in Fig.~\ref{fig:periodogram}.
The most significant signal is found
at a period of $0.369829 \pm 0.0000036$\,d (frequency $\sim$2.70\,d$^{-1}$), which we interpret as the stellar rotation period and, henceforth, denote it by \prot.
The \textit{TESS} light curve phase-folded to \prot\ is shown in Fig~\ref{fig:TESS_VH}. 
We consider the other formally highly significant peak at about $0.1848$\,d,
which is very close to the rotational period reported by \citet{West2015} and \citet{Schoefer2019}, a semi-period of the first because its power is about four times lower and nearly half the value of \prot. The \citet{West2015} period determination was based on MEarth data prior to 2011. Interestingly, an analysis of the SuperWASP light curves, which span several seasons, shows an evolution in the dominant periodicity from $0.1849$\,d before 2011 to $0.3697$\,d after 2013, as illustrated in Figure \ref{fig:SWASP_GLS}.

This evolution of dominant periodicity was previously described in \citet{Basri2018} and \citet{Schoefer2019}. They proposed a geometrical solution to the periodicity shifts. The simplest, solution being two spots 180\,deg apart on the stellar surface. In Sections~\ref{sec:Dis:loc} and~\ref{sec:Dis:mflares} we show evidence for two active regions on opposite hemispheres of GJ~3270. These regions, however, are traced using chromospheric activity indicators that are indicative of faculae or plague features more than spots. While we surmise that the periodicity shifts can arise from any bimodal surface, or near surface, heterogeneity that varies in relative strength over time, the driving mechanism is behind this phenomenon remains unclear. We consider this issue in need of further research because it has strong implications on determining whether a radial velocity (RV) signal is a potential close-in planet or stellar activity.

\citet{Reiners2018} report the $v\sin i$ of GJ~3270 to be 35.3$\pm$3.5~km\,s$^{-1}$ and \citet{Kasseli2018} report a $v\sin i$ value of 37.3$\pm$1.3\,km\,s$^{-1}$. In this work, we adopt the more conservative estimate by \citeauthor{Reiners2018}, but we note that using the \citeauthor{Kasseli2018} values does not appreciably alter the results of this paper\footnote{The $v\sin{i}$ value of 190.3\,\kms\ reported by \citet{Jeffers2018} was incorrect.}. 
Combining the $v\sin i$ with the photometric rotation period, \prot, the value for the stellar radius can be constrained as follows:
\begin{equation}
    R =\frac{P_{rot} v\sin{i}}{2{\pi}\sin{i}} \geq \frac{P_{rot} v\sin{i}}{2{\pi}} \; .
    \label{Eq.:Radv}
\end{equation}
This yields a lower radius limit of
$0.26 \pm 0.05\,R_{\odot}$, which is
consistent with the 0.278$\pm$0.009\,$R_{\odot}$ determined by \citet{Cifuentes2020}. Out of an abundance of caution we doubled the error bars we had initially calculated owing to the youth of GJ~3270. This radius determination lends support to the conclusion that the 0.1848\,d period is a semi-period of the 0.3698\,d period.

Adopting the latter radius allows us to estimate the inclination of the stellar rotation axis, which yields a best estimate of $68\pm 15$\,deg for the inclination and a value of $38\pm 1.2$\,km\,s$^{-1}$ for the equatorial rotation velocity. We present these and other known parameters of GJ~3270 in Table~\ref{tab:properties}.

\begin{table}
\centering
\caption{GJ~3270 basic properties.}
\label{tab:properties}
\begin{tabular}{l c l} 
\hline\hline
\noalign{\smallskip}
Parameters & LSPM~J0417+0849    & Ref.\\
\noalign{\smallskip}
\hline
\noalign{\smallskip}
Karmn\tablefootmark{a} & J04173+088 & Cab16 \\
$\alpha$ (J2000) & 04:17:18.52&  {\it Gaia}\\
$\delta$ (J2000) &+08:49:22.10&{\it Gaia}\\
$d$ [pc]&14.59$\pm0.02$&{\it Gaia}\\
$G$ [mag] & 11.3537$\pm$0.0013 &{\it Gaia}\\
Sp. type & M4.5\,V& PMSU\\
$T_{\rm eff}$ [K] & 3100$\pm$200 & This work\tablefootmark{b} \\
$L$ [$L_{\odot}$]&0.00642$\pm$0.00003&Cif20\\
$R$ [$R_{\odot}$]&0.278$\pm$0.009&Cif20\\
$M_{\star}$ [$M_{\odot}$]&0.269$\pm$0.013&Cif20\\
pEW (\ha) [$\AA$] & $-11.5 \pm 0.015$ &Schf19\\
$P_{\rm rot}$~[d] &0.369829$\pm$0.0000036&This work\\
$v\sin{i}$~[km\,s$^{-1}$] &35.3$\pm$3.5&Rei18\\
$v$~[km\,s$^{-1}$]& 38.0$\pm$1.2 & This work\\
$i$ [deg] & 68$\pm$15&This work\\
$U$ [km\,s$^{-1}$] & $-7.75\pm 4.7$  & CC16 \\
$V$ [km\,s$^{-1}$] & $-27.03 \pm 0.38$  & CC16 \\
$W$ [km\,s$^{-1}$] & $-15.13 \pm 2.57$  & CC16 \\
\noalign{\smallskip}
\hline
\noalign{\smallskip}
\end{tabular}
\footnotesize{{\bf References:} 
{\it Gaia}: \cite{GAIADR2}; 
Cab16: \cite{Caballero2016a}
Cif20: \cite{Cifuentes2020}; 
PMSU: \cite{Hawley1996}; 
Schf19: \cite{Schoefer2019}; 
Rei18: \cite{Reiners2018};  
CC16: \cite{Cortes-Contreras2016}. 
{\bf Notes.} $^{ (a)}$ CARMENES identifier. $^{ (b)}$ Based on original $T_{\rm eff}$ determination by \citet{Cifuentes2020}.
}
\end{table}

\section{Analysis}

\label{sec:Analysis}

In this section we lay out the methods to be used in the analysis of the photometric and spectroscopic data.

\subsection{Flare energy estimation}
\label{sec:Analysis:Photometry}

Our normalized multiband light curves show easily recognizable flare
signatures above the underlying photospheric background, but these light curves lack an absolute calibration because no photometric standard stars were available in our FOV.
To obtain fluxes and luminosities for the flares, we used a PHOENIX model spectrum (see Sect.~\ref{sec:stellarPars}) as an absolute reference for the photospheric spectrum. We obtained band-specific stellar surface fluxes, $f_b$, by
folding the PHOENIX spectrum with the respective filter
transmission curves. Multiplication with the stellar surface area (see Sect.~\ref{sec:stellarPars}) then yielded the band-specific photospheric luminosity, $L_b$, against which the flare is observed.

To study the flare parameters, we set up a light curve model with an exponential form. The free parameters are the flare start time, $t_0$, the peak, $l_p$, and the (exponential) decay time, $\tau$. We also include an offset, which we consider a nuisance parameter. As in particular the {\em TESS} light curves show relatively long integration times per photometric data point, we used   
an oversampled model light curve, which we subsequently binned to the temporal resolution of the respective measurement \citep[e.g.,][]{Kipping2010}. We then obtained best-fit parameters with a $\chi^2$ minimization. The model provided our normalized photometric light curve $l_b (t_i)$ at time $t_i$. We obtained flare luminosities, by

\begin{equation}
    L_{F,b} (t_i) = L_b \cdot l_b (t_i) \; .
    \label{Eq.:flareLumi}
\end{equation}

The total flare energy in the band, $E_b$, is obtained by
integration of luminosity over the flare period. 
Assuming an exponential form for the flare light curve,
the peak luminosity, $L_{\rm peak, b} = l_p \cdot L_b$, and total flare energy are related to the $e$-folding time, $\tau_b$, through
\begin{equation}
    \tau_b = \frac{E_b}{L_{\rm peak, b}} \; .
    \label{Eq.:Tau}
\end{equation}
We estimate that the relative
uncertainty of the total energy amounts to about $10$\,\%,
primarily caused by the systematic error induced by using the synthetic template.

\subsection{Flare model}
\label{sec:Analysis:Flare Model}

As we have simultaneous multiband light curves, an
estimation of the temperature and the size of the flaring region can be attempted. We first degraded
the time resolution of the individual light curves to align their time binning with that of the $V$-band light curve.
To that end, we averaged all $r$, $i$, and $z$-band photometric
data points falling into the respective $V$-band time bin and linearly interpolated the $B$-band light curve. 

In our modeling,
we adopt a single blackbody with a temperature $T_{bb}$ for the flare spectrum. By scaling the flare spectrum with the flare area, $A_f$, and folding with the filter transmission curves, we simulated the response of the different photometric bands. 

For each time bin of the rebinned light curve, we
estimate values for the blackbody flare
temperature, $T_{bb}$, and its area, $A_f$, by
fitting the model to the five available band fluxes.
In the fit, we
gave equal weight in the individual light curves by
assuming a signal-to-noise ratio (S/N) of 100 for all of them. The fits in the individual time bins
are independent, with the exception that
we demand that the blackbody
temperature does not rise after the flare peak.

\subsection{Spectroscopic index definition}
\label{sec:Analysis:Spectroscopy}

We employed the following lines as chromospheric activity indicators: He~{\sc i} D3 $\lambda$5877.2\,\AA\ (henceforth \hei), Na~D2 $\lambda$5891.5~\AA\ (henceforth \sii), Na~D1 $\lambda$5897.5\,\AA\ (henceforth \si), \ha\ $\lambda$6564.6\,\AA, and the \ion{Ca}{II} infrared triplet B line at 8500.4\,\AA\ (henceforth \ca). We focused on the latter component because the \ion{Ca}{II} IRT A \& C lines are closer to the edge of the CCD and subject to greater uncertainty. 
The He~{\sc i} $\lambda$10\,830\,{\AA} triplet lines are heavily affected by telluric OH emission lines and are, therefore, not suitable for our analysis. The Na D lines also show some telluric contamination, which mainly affects the last hour of our observations because GJ~3270 was low on the horizon, but this does not impede our analysis. 

We used the index method as described by \citet{Kurster2003} to quantify the state of the chromospheric indicators. For each line, $L$, index values, $I_{L}$, are calculated for all spectra according to

\begin{equation}
    I_{L} = \frac{\overline{F_{T,L}}}{\frac{1}{2} (\overline{F_{R_{L,1}}} + \overline{F_{R_{L,2}}})} \; .
    \label{Eq.:Index}
\end{equation}

In this equation, $\overline{F_{T,L}}$ denotes the average flux density over a target region, covering the respective line core, and $\overline{F_{R_{L,1}}}$ and $\overline{F_{R_{L,2}}}$ indicate averages over reference regions of pseudo-continuum.
We adopted the same width of 5\,\AA\ for all regions.
Details are given in Table~\ref{TB:WRtable}.

\begin{table} 
    \caption{Vacuum wavelength ranges adopted for index definition}
    \label{TB:WRtable} 
    \centering 
    \begin{tabular}{lccc} 
    \noalign{\smallskip}
        \hline \hline
        \noalign{\smallskip}
        Indicator & Target & Reference 1 & Reference 2 \\
        \noalign{\smallskip}
                  & [\AA] & [\AA] & [\AA] \\ \hline
        \noalign{\smallskip}
        \ha\ Index & 6562--6567 & 6550--6555 & 6570-6575 \\
        \ha\ Broad & 6550--6575 & 6520--6545 & 6580-6605 \\
        \ha\ BWI & 6558--6563 & 6540--6545 & 6580-6585 \\
        \ha\ RWI & 6566--6571 & 6540--6545 & 6580-6585 \\
        \ha\ BWI-e & 6556--6561 & 6540--6545 & 6580-6585 \\
        \ha\ RWI-e & 6568--6573 & 6540--6545 & 6580-6585 \\
        \hei\  & 5875--5880 & 5869--5874 & 5905-5910 \\
        \sii\  & 5889--5894 & 5869--5874 & 5905-5910 \\
        \si\  & 5896--5901 & 5869--5874 & 5905-5910 \\
        \ca\ & 8497.5--8502.5 & 8490--8495 & 8505-8510 \\
        \hline 
    \end{tabular} 
\end{table} 

The relatively broad target regions account for the strong rotational line broadening and additional broadening of the chromospheric line
cores during flares.
Uncertainties on the index values were obtained by error propagation as follows:

\begin{equation}
    \sigma_I = I \times \sqrt{\left (\frac{\delta \overline{F_{T}}}{\overline{F_{T}}}\right)^2+
    \left (\frac{\delta\overline{F_{R_{1}}}}{\overline{F_{R_{1}}}+\overline{F_{R_{2}}}}\right)^2+
    \left (\frac{\delta \overline{F_{R_{2}}}}{\overline{F_{R_{2}}}+\overline{F_{R_{1}}}}\right)^2} \; ,
    \label{Eq.:Indexerror}
\end{equation}

\noindent where $\delta \overline{F_{T}}$, $\delta \overline{F_{R_{1}}}$, and $\delta \overline{F_{R_{2}}}$ denote the uncertainties of the respective mean, and the line index, $L$, was dropped for readability. 
All index values and uncertainties are listed in Table ~\ref{indextable}.

To better study the relation between the activity indices, we created a relative index (henceforth $I/I_r$) for every line, $L$, such that
\begin{equation}
I_{r,L} (t_i) = \frac{I_L (t_i)}{I_{L} (t_{\rm low})},
\end{equation}
where $t_{\rm low}$ denotes the minimum activity state, corresponding to the spectrum with the lowest observed \ha\ index value (observation no. 13). This was also the case for all the other activity indicators except for \hei, which had the first exposure, of our observation period, as its lowest value.

\section{Results}
\label{sec:Results}

We applied the methods discussed in Sect.~\ref{sec:Analysis} to the photometric (Sect.~\ref{sec:Results:photometry}) and spectroscopic (Sect.~\ref{sec:results:spectroscopy}) data. 
In Sect.~\ref{sec:SvP} we compared these results with emphasis on timing and energy differences of the effect of the flare in the photometric and spectroscopic data.

\subsection{Photometry}
\label{sec:Results:photometry}

We present the results from analyzing the photometric data from SNO and MuSCAT2 (Sect.~\ref{sec:R_Phot}) and \textit{TESS} (Sect.~\ref{sec:TESS}). 
We emphasize the determination of the energy, peak luminosity, and $e$-folding decay time of the flares. 
This allows us to determine that the flares that we observed on 15 December 2018 do not differ greatly when compared to the characteristics of the flares \textit{TESS} observed the month prior.

\subsubsection{SNO and MuSCAT2}
\label{sec:R_Phot}

\begin{figure}
    \centering
    \includegraphics[width=0.5\textwidth, angle=0]{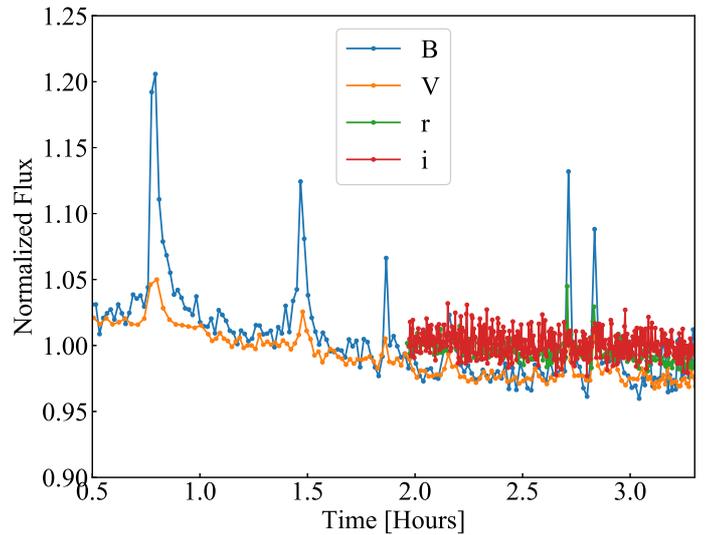}
    \caption{Minor flaring in $B$, $V$, $r$, and $i$ bands that occurred at the beginning of the observation run.}
    \label{fig:miniflares}
\end{figure}

The light curves in Fig.~\ref{fig:muscatsno} show two prominent flare-like events at relative times close to 3.6\,h (henceforth flare~1) and 4.2\,h (henceforth flare~2), of which the latter
shows higher peak flux in all bands. 
At least five weaker flares were observed in the $B$ and $V$ bands, primarily early in the observing run (Fig.~\ref{fig:miniflares}).

\begin{figure}
    \centering
    \includegraphics[width=0.5\textwidth, angle=0]{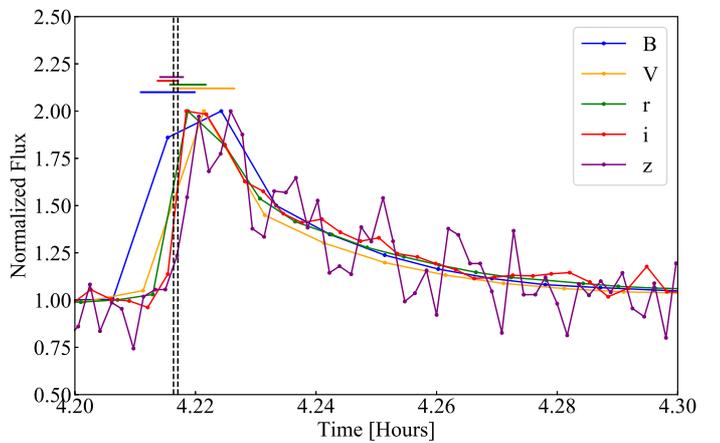}
    \caption{Multiband light curve of flare~2 normalized by peak flare flux. The horizontal bars indicate the integration time of the flare onset observation for each band. Because of the low S/N of the $z$-band observations, we extended the possible onset time of the flare by an additional exposure. The dashed lines represent the onset window for flare 2 that would satisfy the onset conditions of all the photometric bands.}
    \label{fig:Fcurves}
\end{figure}

The different exposure times and relative offsets in cadence complicate the determination of the instant of flare onset.  
We normalized the flare light curves by the peak flux (Fig.~\ref{fig:Fcurves}). This allowed us to identify a 2.6~s window of overlap between the first bins, which consider to show elevated flux due to flaring. Therefore we determined that the onset of flare 2 occurred
at UT 15-12-2018 23:47:33 (4.2167\,h into the observations, JD~2458468.49135 $\pm$1.3\,s), which is consistent with simultaneous onset in all light curves. While we consider this strong evidence that a simultaneous onset occurred, it does not eliminate the possibility of a delayed onset. We are able to constrain any such delay to a maximum of 25\,s between onset in $B$ and onset in $z$, which would still satisfy the observed light curves.
 
We applied the methods described in Sect.~\ref{sec:Analysis:Photometry} to determine the band energy, peak luminosity, and $e$-folding decay time. We present these values in Table~\ref{TB:F12P}. The results for the $z$ band in flare~1 remain insignificant and are, therefore, not listed in the table. The $e$-folding decay times are presented in Table~\ref{TB:F12P}.

Flaring is generally more pronounced in the bluer bands, which is a consequence of the higher temperature flare spectra contrasting against the cooler stellar spectra \citep{McMillan1991,Rockenfeller2005}. This behavior is also 
exhibited by flare 2 as shown in Fig.~\ref{fig:muscatsno}. This effect is so pronounced that flare~1 becomes essentially undetectable in the $z$ band. From our best-fit model (see Sect.~\ref{sec:Analysis:Flare Model}) we determine the peak temperature of flare 2 to be $\sim$10,000$\pm$1500\,K, covering an area of (1.35$\pm$0.39)$\times10^{19}$\,cm$^2$. Using the same procedure to calculate the total bolometric luminosity assuming the flare is a blackbody as in \citet{Gunther2020} and \citet{Shibayama2013}, we estimate the total energy to be $(6.3 \pm 3.2)\times10^{32}$\,erg, which is consistent with the sum of $B$,$V$,$R$,$i$, and $z$ band energies of $3.6\times 10^{32}$\,erg. This puts flare 2 close to the superflare regime of flares with bolometric energy greater than about $10^{33}$\,erg \citep{Shibayama2013}. 

However, while the model fits well to the peak and early decay phase, it does not reproduce the onset or late decay phase. For the onset we suspect that the issue lies in the aforementioned differences in cadence and exposure and the swift development of the flare. For the late decay phase the parameters become degenerate as the flare-affected region cools and, therefore, are no longer reliable. In our energy calculation we omit these degenerate values but nevertheless we assign a conservative 20\,\% error on the energy estimate.

\begin{table} 
    \caption{Parameters of flares~1 and 2\tablefootmark{a}.}\label{TB:F12P} 
    \centering 
    \begin{tabular}{l c c c c} 
        \hline \hline 
\noalign{\smallskip}
       Band & $E_b$  & $L_{{\rm peak},b}$ & $\tau_b$ & $\frac{E_b}{\Delta \lambda_{b}}$\\
            & [$10^{31}$\,erg]  & [$10^{29}$\,erg\,s$^{-1}$]  & [s]& [$10^{29}$\,erg\,$\AA^{-1}$]  \\ \noalign{\smallskip}
\hline
\noalign{\smallskip}
       \multicolumn{5}{c}{\em Flare 1} \\
       $B$ & 3.19 & 3.63 & 88 & 0.41\\  
       $V$ & 1.86 & 1.98 & 99&0.19  \\  
       $r$ & 1.56 & 1.34 & 116&0.12 \\  
       $i$ & 1.18 & 0.68 & 85&0.09 \\ 
       $z$ & ... & ... & ...& ... \\  
\noalign{\smallskip}
       \hline
\noalign{\smallskip}
       \multicolumn{5}{c}{\em Flare 2} \\
       $B$ & 12.95 & 19.79 & 65&1.66 \\  
       $V$ & 7.14 & 11.46 & 62&0.72 \\  
       $r$ & 6.12 & 7.17 & 85&0.48 \\  
       $i$ & 3.65 & 3.87 & 94&0.28 \\  
       $z$ & 5.89 & 7.19 & 82&0.22 \\  
\noalign{\smallskip}
        \hline 
\noalign{\smallskip}
    \end{tabular} 
    \footnotesize{ {\bf Note:} $^{ (a)}$ $\Delta \lambda$ in the fourth column denotes FWHM of the filter. Values given in Section \ref{sec:GBP}.}
\end{table}

\subsubsection{\em TESS}
\label{sec:TESS}

We carried out a search for flares in the \textit{TESS} light curve.
To that end, we classified any photometric excursion as a flare if it peaked more than 3.6$\sigma$ above the noise (determined by comparing lowest detectable injected flare to the noise background) and consisted of more than three consecutive data points.
As the data set is inherently skewed owing to the presence of flares, we used the robust median deviation about the median to estimate the standard deviation of noise \citep[e.g.,][]{Rousseeuw1993}. 
With our criterion, we identified 22 flare events in the 26 day observational period. Our detection threshold for these flare events is \textasciitilde10$^{31}$erg.
By visual inspection, we determined that 15 of these are isolated events, with a single distinct peak followed by a decay. The remaining seven events were part of three different features consisting of multiple peaks. This occurs when a new event begins prior to the end of the decay phase of the previous event. Whether the events are physically related or are aligned by chance is unclear.
Of these seven events, three were rejected from processing. One of these was rejected because if it been an isolated event, it would not have met the 3.6$\sigma$ criterion. Two others were rejected as their profiles were not able to be fit with an exponential owing to an unusually long decay time or confounding behavior of the continuum. Likewise, one isolated event was rejected for the same issue.
The parameters of the remaining flares are given in Table~\ref{TFP}.

\begin{table} 
    \caption{Parameters of \textit{TESS} flares.}
    \label{TFP} 
    \centering 
    \begin{tabular}{l c c c c} 
        \hline 
        \hline 
        \noalign{\smallskip}
       Type & $E_T$ & $L_{{\rm peak},T}$ & $\tau_T$ & Onset time\tablefootmark{a}\\
        & [$10^{31}$\,erg] & [$10^{29}$\,erg\,s$^{-1}$] & [s] & [d] \\
\noalign{\smallskip}
       \hline
\noalign{\smallskip}
       S & 6.70 & 6.32 & 106 & 1.2917  \\  
       S & 33.68 & 4.63 & 721 & 3.9806\\  
       S & 4.83 & 7.10 & 68 & 6.1001\\  
       S & 58.25 & 7.53 & 774 & 7.7682\\  
       S & 6.96 & 3.12 & 223 & 7.8668 \\  
       S & 10.22 & 5.03 & 203 & 9.4015 \\  
       S & 11.57 & 3.49 & 332 &13.9043\\  
       S & 38.07 & 21.28 & 179 &15.1598\\  
       S & 30.72 & 26.69 & 115 & 17.4195 \\  
       S & 15.02 & 3.02 & 498 & 21.0736 \\  
       S & 12.34 & 1.96 & 628 & 21.6111\\  
       S & 3.01 & 2.49 & 121 & 22.6402\\  
       S & 8.15 & 3.88 & 210 & 24.0666\\  
       S & 24.01 & 13.55 & 177 &24.2305\\  
\noalign{\smallskip}
       \hline
\noalign{\smallskip}
       M & 14.22 & 14.79 & 96&2.6334 \\  
       M & 18.85 & 8.32 & 222&2.6403 \\  
       M & 10.69 & 2.36 & 452&3.6528 \\  
       M & 12.74 & 13.27 & 96&11.0556 \\  
 
\noalign{\smallskip}
        \hline 
\noalign{\smallskip}
    \end{tabular} 
    \footnotesize{ {\bf Notes.} $^{(a)}$ Time given in days since beginning of \textit{TESS} observations, sector 5 (JD 2458437.997). S indicates the flare was an isolated event, whereas M indicates the flare was part of a MFE event. Values represent the output of the \textit{TESS} band only.}
\end{table} 

For individual flare events (IFEs) the peak luminosities, energies, and $e$-folding decay times were calculated in the same manner as that described in Sect.~\ref{sec:Analysis:Photometry} for the SNO and MuSCAT2 data. The flares in the multi-flare events (MFEs) were a bit more complicated in that the decay curves of the individual events were disrupted by the curve of the following event. To estimate the energy, we fit an exponential decay curve based on the peak flux and the data points that existed prior to the next flare occurrence. The same procedure was done for the second and third flare in the sequence, only now subtracting the curve of the prior events. This allowed us to integrate under the curves and get an approximation for the energy output in each flare as if they were individual events. The values of these calculations are given in Table~\ref{TFP}.

\begin{figure}
    \centering
    \includegraphics[width=0.49\textwidth, angle=0]{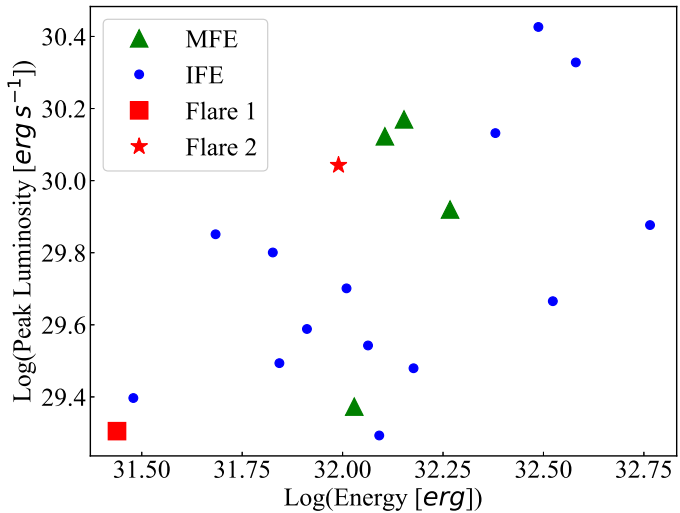}
    \caption{Flare peak flux as a function of total energy
    for flares~1 and 2 (red square and star, respectively) as well as multiple (green triangles: MFE) and individual flare (blue circles: IFE) events observed by \textit{TESS}.
    }
    \label{fig:Tlog}
\end{figure}

In Fig.~\ref{fig:Tlog}, we show the flare peak luminosity as a function of flare energy.
In comparing the data on the flares observed by \textit{TESS} to those of flares 1 and 2, we opted to use the values in the $i$ band and $r$ band, added together.
We opted to not use the $z$ band as well because only flare~2 was detected in this band.

Isolated and MFEs behave similarly in this diagram, albeit the number of MFEs remain low.
While others have identified these MFEs \citep{Tsang2012,Vida2019b,Gunther2020}, also referred to as outbursts, there is to our knowledge no direct comparison of the parameters of single versus multiple events. Taken in total, however, the data show a statistically significant relation between peak luminosity and energy (p-value of 4.96$\times10^{-6}$).

\subsection{Spectroscopy}
\label{sec:results:spectroscopy}

We present the results of analyzing the CARMENES spectroscopic data. The descriptions of the indicators used can be found in Sect.~\ref{sec:Analysis:Spectroscopy}. First, we present the response of the chromospheric activity indicators (Sect.~\ref{sec:indextimeseries}). These give broad outlines of the activity state of the star at any given time during the observations. Following this, we use the same methodology to examine the asymmetry of the \ha\ line (Sect.~\ref{sec:wingindex}). In Sect.~\ref{sec:velocity} we show that the blue asymmetry in \ha\ is caused by blue-shifted emission feature that is strongly correlated with those of the other activity indicators.

\subsubsection{Chromospheric index time series}
\label{sec:indextimeseries}

\begin{figure*}
    \includegraphics[width=0.5\textwidth, angle=0]{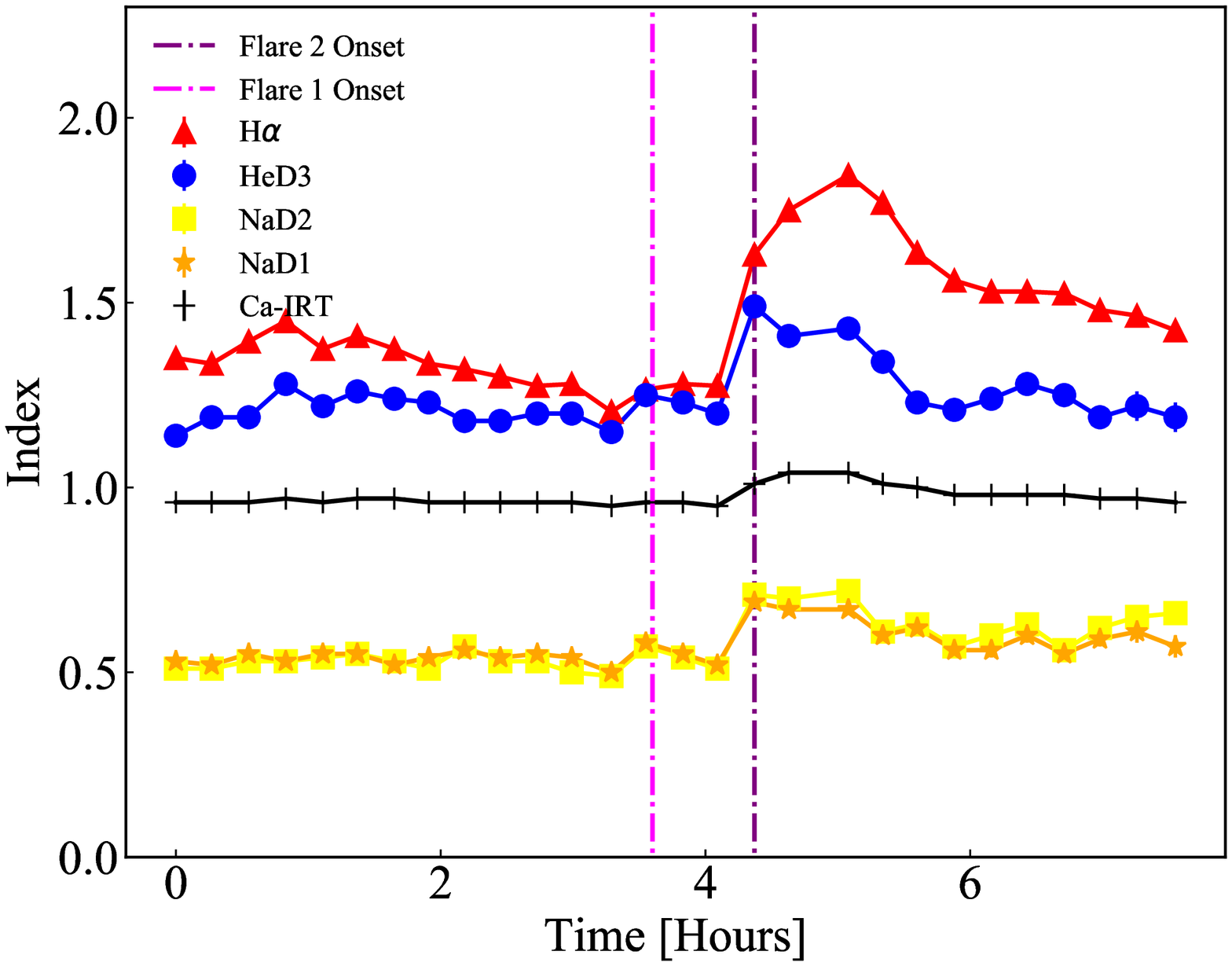}
    \includegraphics[width=0.5\textwidth, angle=0]{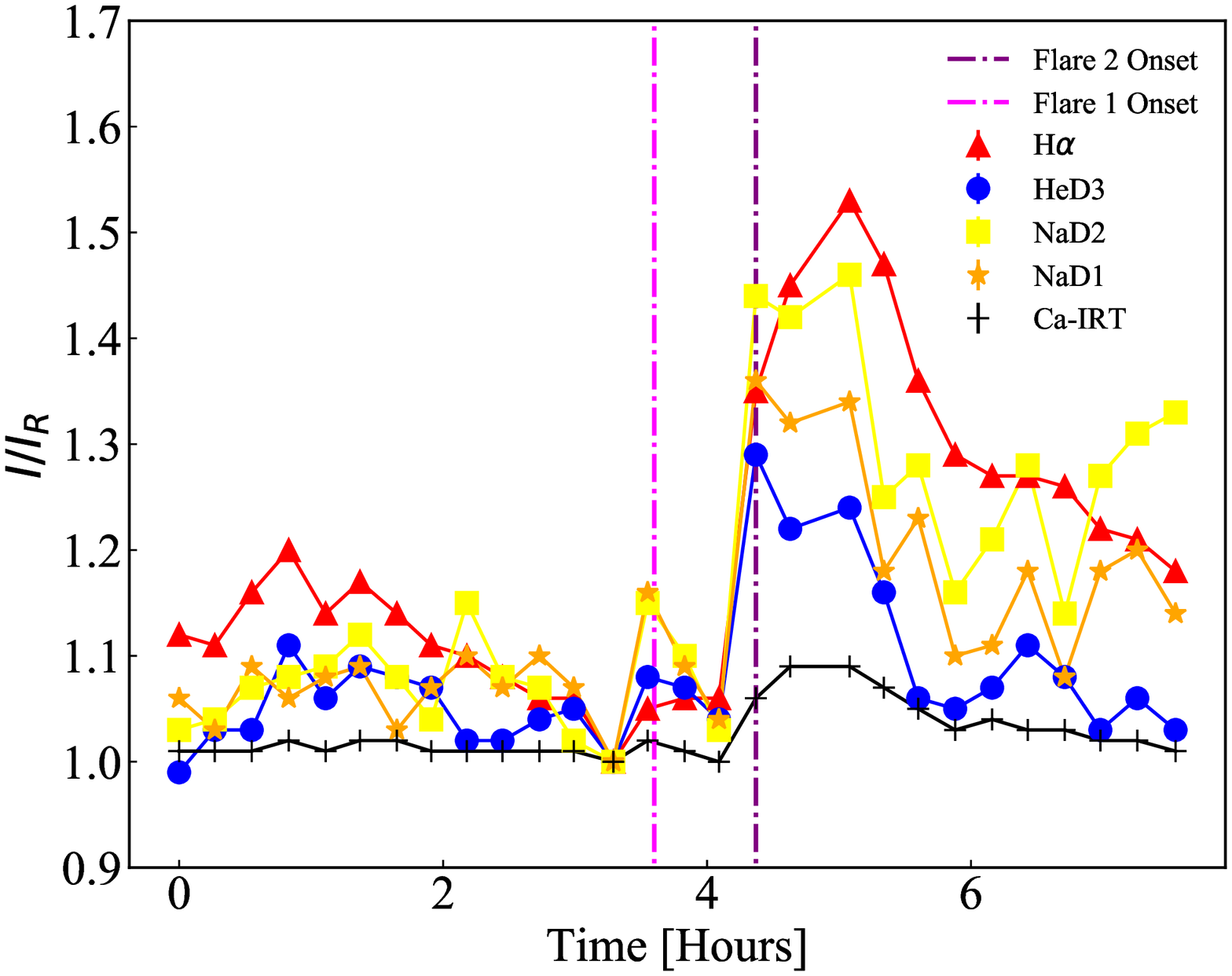}
    \caption{\emph{Left:} Index for \ha\  (red triangles), \hei\  (blue circles), \sii\  (yellow squares), \si (orange stars), and \ca\  (black crosses) for all observations. Flare 2 occurs at 4.3 h.  
    \ha\ is scaled down by factor of 2.5.
    \emph{Right:} $I/I_r$ for  \ha \ (red triangles), \hei\  (blue circles), \sii\  (yellow squares), \si\  (orange stars), and \ca\  (black crosses) for all observations. Vertical lines in both plots indicate onset times of flares 1 and 2. }
    \label{fig:Index}
\end{figure*}

In Fig.~\ref{fig:Index}, we show the time evolution of chromospheric indices.
Prior to flare~1 (at about 3.6\,h), all the activity indices remained  roughly constant with the exception of that of \ha, which showed a decline. 
Flare 1 did not elicit a strong response from any activity indicators. 
If not for the photometric signature, flare 1 would have remained undetected. 
Interestingly, the sodium indices seemed to respond the most, in relation to the response of other indicators, to flare 1 (Fig.~\ref{fig:Index}).  
In contrast, all indicators showed a pronounced flare signature that is consistent with the timing of photometric flare~2, with a characteristic fast rise and subsequent longer decay phase \citep[][]{Fuhrmeister2018, Reiners2008, Honda2018, Schmidt2019}. 

The most prominent flare~2 response was observed in the \ha\ line, for which the rise to the peak index value lasted about 45\,min. The following decay phase slowed after another 45\,min essentially developing a plateau, followed by a moderate linear decay lasting beyond the end of the observations. In comparison to \ha, \ca\ reacted the least but reached its peak faster. However, similar to \ha\, this flare did not fully return to its pre-flare values before the end of observations.

The \hei\ and Na~{\sc i} indices showed similar temporal behavior, marked by a swift (unresolved by 15\,min cadence) rise phase followed by a 45\,min plateau. 
After this plateau, the \hei\ and the sodium line indices showed a decay back to pre-flare levels, which appears to proceed the most rapidly in the \hei\ line index. 
Close to the 6 h mark, the lines showed another enhancement in activity, which coincided with the plateau-like feature seen in the \ha\ index. The pattern of \ca\ decaying slower than \hei\ has been previously noted with other M dwarf flares \citep{Fuhrmeister2008,Fuhrmeister2011}.

The \si\ and \sii\ indices showed an apparent increase in activity level after 6.5 h into the observations.
This period coincided with increased telluric contamination of the sodium lines (Fig.~\ref{fig:heatmaps}).

\subsubsection{\ha\ wing indices}
\label{sec:wingindex}

\begin{figure}
        \centering
        \includegraphics[width=0.5\textwidth, angle=0]{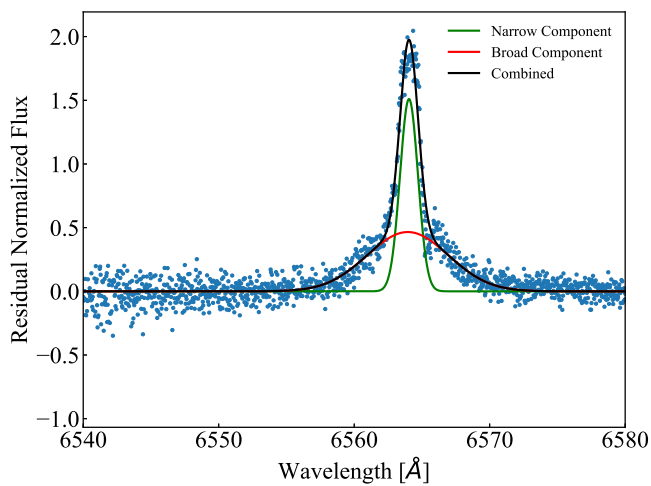}
        \caption{\ha\ double Gaussian fit for the onset of flare 2. The narrow component (green line) has a shift of $-25.12$\,km\,s$^{-1}$ and the broad component (red line) has a shift of $-29.68$\,km\,s$^{-1}$ from the line core. The combined fit is given by the solid black line. }
        \label{fig:DG_Flare_Onset}
    \end{figure}

Flare~2 had effects beyond the core of the \ha\ line profile with an increase in flux being detected from {$-10$~\AA}\ to $+5$~\AA. We show this in Fig.~\ref{fig:DG_Flare_Onset} with line fit parameters given in Table~\ref{TB:DGtable}. Moreover, we quantified the effects in the \ha\ wings, including asymmetries, by measuring an index on both sides of the core index: the red wing index (RWI) and blue wing index (BWI).

\begin{figure}
    \centering
    \includegraphics[width=0.5\textwidth, angle=0]{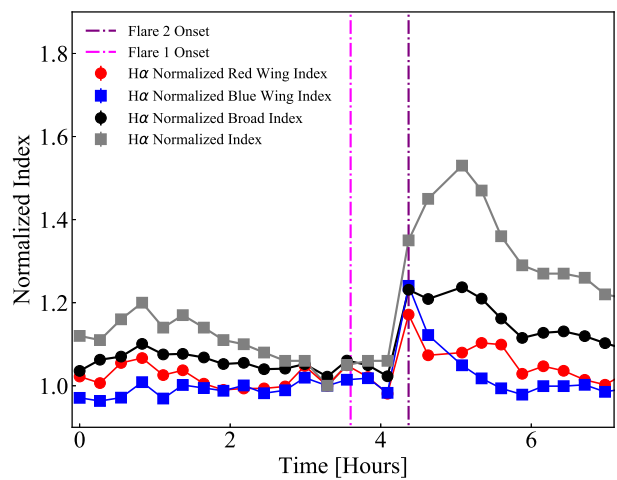}
    \caption{\ha\ red (RWI) and blue (BWI) wing indices. \ha\ $I/I_r$ and broad index are shown in gray and black, respectively. The vertical lines indicate onset times of flares 1 and 2.}
    \label{fig:WingIndex}
\end{figure}

Additionally, we took the index of a broad region that included the line core (Broad Index). The definitions of these indices can be found in Table~\ref{TB:WRtable}. We plot these new indices along with the original \ha\ index in Fig.~\ref{fig:WingIndex}. 

\begin{table} 
    \caption{Parameters of double Gaussian fit.}
    \label{TB:DGtable} 
    \centering 
    \begin{tabular}{lcc}
    \hline \hline
     \noalign{\smallskip} 
    Parameter & Narrow component & Broad component \\
     \noalign{\smallskip} 
    \hline
     \noalign{\smallskip} 
    Amp. [\AA] & 1.51 & 0.47 \\
    $\delta$ Amp. [\AA] & 0.03 & 0.03 \\
    $\mu^a$ [\AA] &6564.05\,(-25.12)& 6563.95\,(-29.68)\\
    $\delta\,\mu^a$ [\AA] &0.01\,(0.46)&0.01\,(0.46)\\
    $\sigma$ [\AA] &0.66&3.0\\
    $\delta\, \sigma$ [\AA] & 0.02&0.13 \\
    \hline
    \end{tabular}
    \tablefoot{$^a$ Figures in parenthesis denote the Doppler shift in km\,s$^{-1}$.}
\end{table} 

At the beginning of the observations, the RWI was enhanced over that of the BWI, but by 2 h into the observations the two indices equalized. Neither index reacted to the onset of flare 1. The BWI showed a sharp rise at the onset time of flare 2 followed by an exponential decay with an $e$-folding time of about $30$\,min. The RWI also showed a rapid increase at the time of the onset of flare 2 followed by a decay. This decay was interrupted, however, and a secondary rise was detected shortly after 5\,h into the observations. The RWI remained elevated over that of the BWI for the remainder of observations. This indicates that the BWI was primarily affected by the events around the flare onset.

\begin{figure}
    \centering
    \includegraphics[width=0.5\textwidth, angle=0]{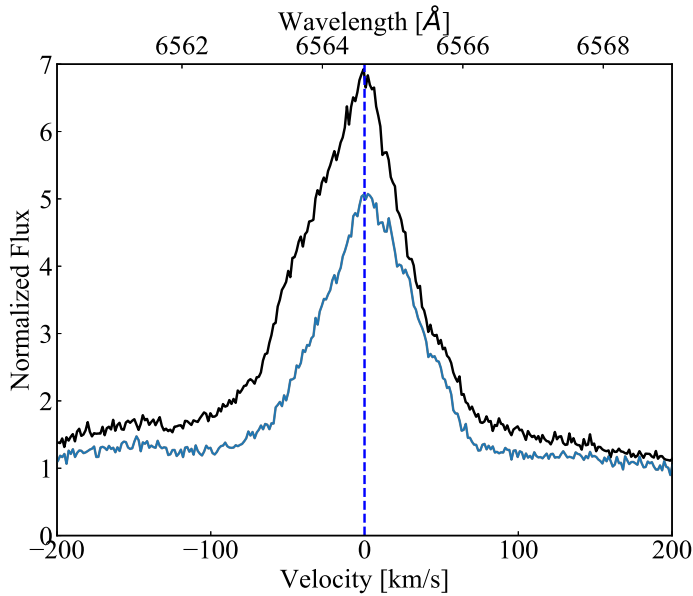}
    \caption{\ha\ at flare onset (solid black, observation 17) and at activity minimum (solid gray, observation 13). The dashed vertical line indicates the rest wavelength.}
    \label{fig:Flare_Onset_H}
\end{figure}

Initially the flare onset appeared in the \ha\ line as a large asymmetry on the blue wing of the line. When subtracting the minimum activity spectrum (Fig.~\ref{fig:Flare_Onset_H}) this blue asymmetry is the summation of two features: a narrow and broad component (Fig.~\ref{fig:DG_Flare_Onset}). We fit a double Gaussian profile to these components, the result of which is given in Table~\ref{TB:DGtable}. 

For the broad component, the fit yielded a shift of $-29.7$\,km\,s$^{-1}$ from the line core and a narrow component shifted by $-25.1$\,km\,s$^{-1}$. This broad component was only strong enough to be fit with a Gaussian at the flare onset, but indices placed even further from the line core than the BWI and RWI  indicate that this component lasted for a total of 30\,min. This also suggests that the narrow component was limited to the 7\,{\AA} nearest the line core as it persisted in the BWI for much longer. The red extreme index did not show as much of an enhancement after 5 h as the RWI, suggesting that the asymmetry detected by the RWI was also confined to within 7\,{\AA} of the \ha\ line core. 

The narrow component, represented by the BWI, persisted for at least 90\,min while decreasing in strength. 
This can be seen in the activity minimum subtracted spectrum as the narrow component decreasing in amplitude and shifting toward the line core.

\subsubsection{Doppler shifted emission}
\label{sec:velocity}

\begin{figure}
    \centering
    \includegraphics[width=0.5\textwidth, angle=0]{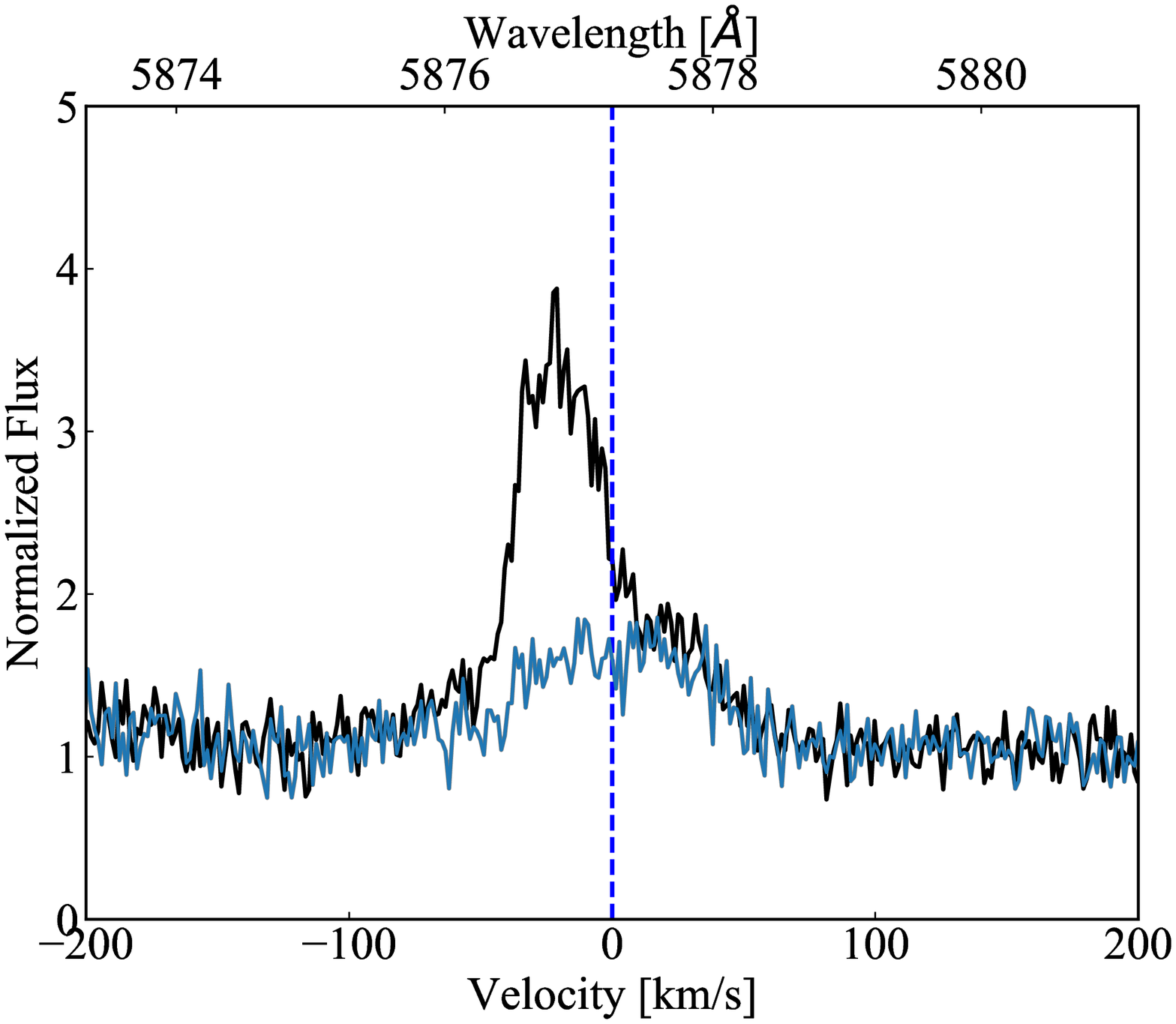}
    \includegraphics[width=0.5\textwidth, angle=0]{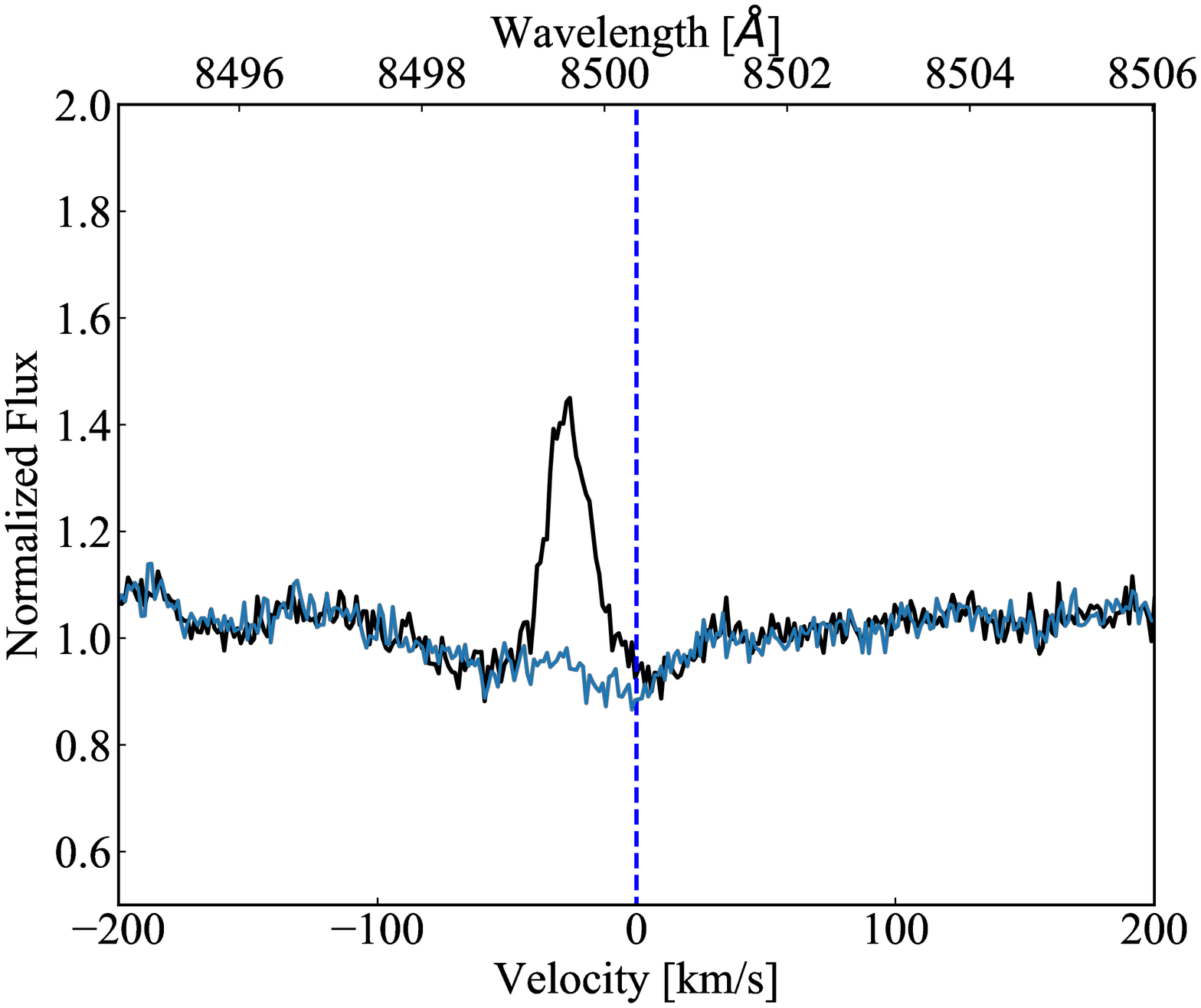}
    \caption{Flare emission from the \hei\ (top) and \ca\ lines (bottom) during flare onset (solid black, observation 17), activity minimum spectra (solid gray, observation 13). The dashed vertical line indicates the rest wavelength.}
    \label{fig:Flare_Onset_D3CA}
\end{figure}

\begin{figure}
    \begin{center}
    \includegraphics[width=0.4\textwidth]{{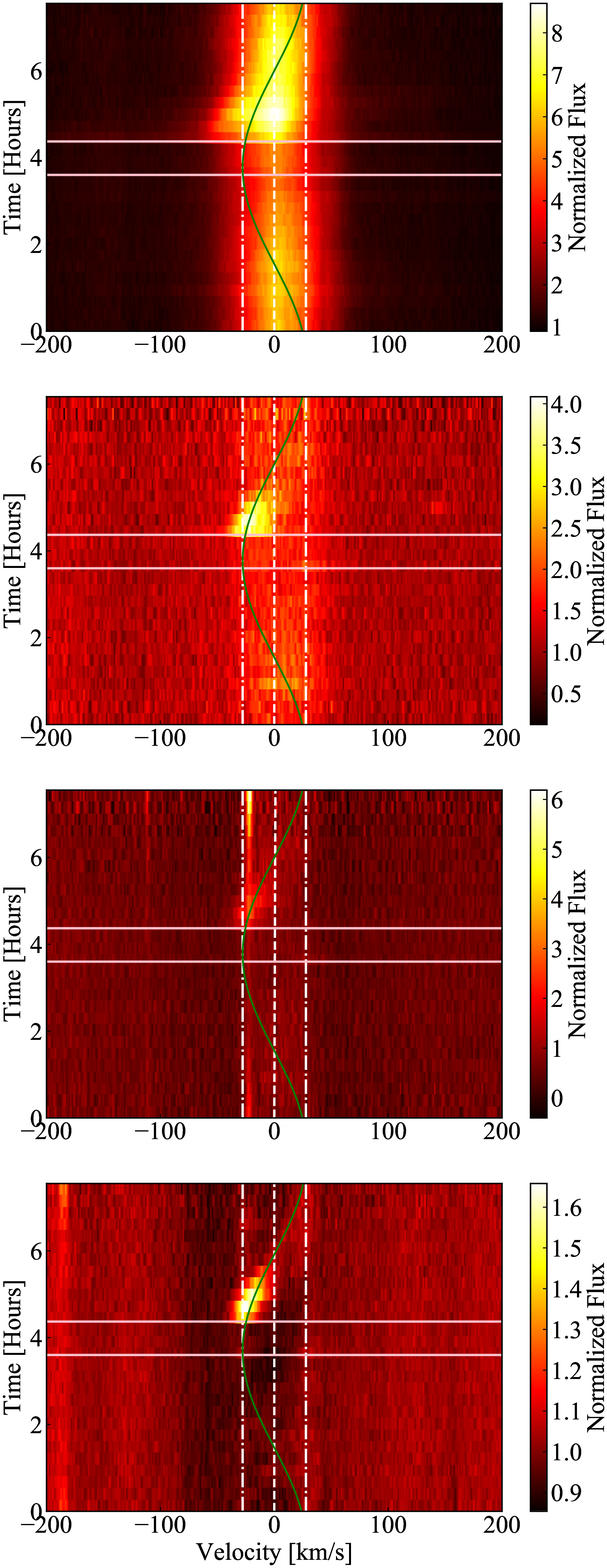}}
    \end{center}
    \caption{From top to bottom: \ha, \hei, \ca, and \sii\ flux density evolution during our observing run.
    The central dashed line indicates the nominal rest-frame wavelength and the outer two dash-dotted lines denote the maximum Doppler shift for a corotating object at 43\,deg latitude (28.6\,\kms). The green line represents the expected Doppler shift of such an object if it were to emerge onto the blue limb of the disk 3.72 h into the observation run. Horizontal solid white lines indicate onset times of flares 1 and 2.}
    \label{fig:heatmaps}
\end{figure}

The initial displacement and subsequent shift toward the line core seen in \ha\ can also be observed in the other activity indicator line profiles (Fig.~\ref{fig:Flare_Onset_D3CA}). To more clearly illustrate this, we compiled all of the available spectra for each indicator into a series of ``heatmaps'' (Fig.~\ref{fig:heatmaps}). These heatmaps show the temporal evolution of the normalized flux density for these lines. The clearest example of the shift is given by the \ca\ line because it has the highest S/N. Notably, none of the activity indicator lines have Doppler displacements that exceed the projected rotational velocity of the star. In this picture, the shift in \ha\ is actually the least obvious owing to the preexisting emission in the line core. 

\begin{figure}
        \centering
        \includegraphics[width=0.5\textwidth, angle=0]{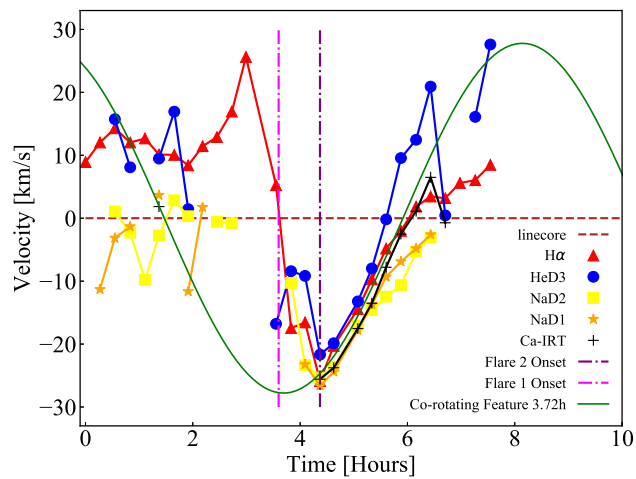}
        \caption{Doppler shift of excess flare emission of the line cores.
        The missing data points indicate that a Gaussian fit was impossible, which causes the gap at about 3\,h, where the minimum activity spectrum is located.
        All of the activity indicators shift to the blue {before} the flare onset (noted by vertical dotted line). Green line represents the expected Doppler shift of such an object, at 43\,deg latitude, if it were to emerge onto the blue limb of the disk 3.72\,h into the observation run.
        }
        \label{fig:Velocity}
\end{figure}

By subtracting the minimum activity spectrum, we can generate a set of residual spectra in which deviations from the quasi-quiescent state can be analyzed. To these residual spectra we fit a Gaussian profile, thereby determining the Doppler displacement from the line core (Fig.~\ref{fig:Velocity}).

All of the Doppler shift values at the onset of flare 2 are between $-27$\,\kms and $-21$\,\kms. The flare onset values of the \ha\ single Gaussian fit are comparable to the results of the narrow component in the double Gaussian fit.
    
Prior to flare~1, the Gaussian fits to the \ha\ and \hei\ residual spectra show predominantly redshifted values, whereas the sodium residuals are either neutral or blueshifted. We attribute these, along with the
sporadic shifts exhibited by the sodium lines, to the low amplitude of the signals, telluric interference, and systematic uncertainty caused by the selection of the minimum activity spectrum.

As the onset of flare~1 approached, the \ha\ residuals became increasingly redshifted. At the onset of flare~1 the \ha\ shift decreased almost to neutrality and \hei\ appeared blueshifted. After the onset of flare~1 but prior to flare~2 \ha, \hei, and the sodium lines all appeared blueshifted with a shift of between $-20$ and $-10$\,\kms. 
After the onset of flare~2 the shifts of all chromospheric residuals were strongly correlated. 
Over the subsequent $\sim$2\,h, the residuals shifted toward longer wavelengths.
Formally, the \hei\ residual emission was the first to return to its rest wavelength and then became increasingly redshifted. 
Shortly after followed the shifts of the residual \ha\ and \ca\ line cores. 
For the sodium lines the return to neutrality could not be observed, probably because of telluric interference.

\subsection{Spectroscopy versus photometry}
\label{sec:SvP}

\begin{table} 
    \caption{Indicator energies for flare 2.}\label{table_lum} 
    \centering 
    \begin{tabular}{lcccc} 
        \hline \hline 
        \noalign{\smallskip}
       Indicator & Band L & Peak L &  Energy &$\tau$\tablefootmark{a} \\ 
       \noalign{\smallskip}
               & [$10^{27}$~erg\,s$^{-1}$] & [$10^{26}$~erg\,s$^{-1}$] & [$10^{29}$~erg]&[s] \\
        \hline
        \noalign{\smallskip}
       \ha\ Index & 3.57 & 45.55 & 136.33 & 2993 \\  
       \ha\ Broad & 17.41 & 50.12 & 371.98 & ... \\  
       \ha\ BW& 3.70 & 9.83 & 11.68 & 1188 \\  
       \ha\ RW & 3.62 & 6.27 & 23.20 & 3700 \\  
       \ha\ EBW & 3.50 & 3.86 & 1.76 & 458 \\  
       \ha\ ERW & 3.49 & 2.65 & 5.40 & 2038 \\  
       \hei\ & 1.07 & 3.61 & 16.96 & 4697 \\  
       \si\ & 0.42 & 7.70 & 4.93 & 6400 \\  
       \sii\ & 0.35 & 7.93 & 5.80 & 7304 \\  
       \ca\ & 6.54 & 5.90 & 34.67 & 5878 \\  
        \hline 
    \end{tabular} 
    \footnotesize{ {\bf Notes.} $^{ (a)}$ $e$-folding decay time (see Sect.~\ref{sec:Analysis:Photometry}), \ha\ broad component does not have a decay value as it does not exhibit an exponential-like decay profile.}
\end{table}

In Fig.~\ref{fig:TESS_VH} we juxtapose the phase-folded \textit{TESS} light curve (with flare events subtracted) with the SNO $V$ band light curve and the \ha\ index time series; the latter two are scaled. The \ha\ index data shows a decline prior to flare~2.

Similar behavior is exhibited by our $B$, $V$, $r$ band light curves, but it is not detectable in the $i$ and $z$ bands, possibly owing to poorer S/N (Fig.~\ref{fig:muscatsno}).
Although the \textit{TESS} light curve was not taken simultaneously, the narrow gap of only about four days, combined with the relative stability of the rotational signal during the TESS observing window, suggests that the source of the decline in Fig.~\ref{fig:TESS_VH} is rotational variability. This rotational variability is likely caused by corotating active regions, which is also consistent with the effect being more pronounced at shorter wavelengths. 
This implies that the modulation of the \ha\ index prior to flare~2 is also, primarily, driven by rotational effects. 

The decay times of the activity indicators for flare 2 are, at a minimum, two times longer than those observed in the photometric bands. 
The $e$-folding duration of \ha, for instance, is $\sim$3000\,s, whereas the longest $e$-folding duration of a photometric indicator is 1320\,s and nearly exponential in nature. 
The decay phase of \ha\ appears to be complex with an initial exponential decay followed by a plateau and a later linear decay that lasts until the end of observations. 
Since the non-exponential decay phase of \ha\ occurred over a period of time when the photometric flare had already returned to quiescent level, the transition between the phases in the activity indicators is likely due to phenomena that affected the chromosphere but not the photosphere. 

Table~\ref{table_lum} shows the energies involved for each activity indicator used in this study. \ha, one of the principal cooling lines of the chromosphere, puts out an energy equivalent to $\sim$22\,\% of the $r$ band. However it takes \ha\ 30 times as long to emit that energy. This contrast in the rate of output between \ha\ and the $r$ band is clear when looking at the peak luminosity. The photometric value is 148 times larger than the \ha\ spectroscopic value. The other spectroscopic indicators have similar ratios to the photometric bands in which they occur. The activity indicators (\hei\ and the sodium lines) in the $V$ band are slightly more contrasted (2.5\,\% of the energy output of the $V$ band) as a consequence of the higher temperature of the flare material in comparison to the photospheric background.

\section{Discussion}
\label{sec:discussion}

We present our interpretation and extrapolation of the data above. In Sect.~\ref{sec:Dis:flares} we compare the flares observed with SNO and \textit{TESS} to other flaring M dwarfs. In Sect.~\ref{sec:Dis:loc} we interpret the Doppler shifts of the activity indicators as evidence for a corotating feature and isolate the location of it to a latitude of43\,deg and a longitude of --70\,deg. In Sect.~\ref{sec:Dis:mflares} we look at the beginning of the observations when sustained minor flaring is associated with a persistent redshift in the \ha\ line. We find that the data are consistent with an active region, separate from that of flare 2, which was in the process of rotating off the observable disk at that time. In Sect.~\ref{sec:Dis:Rotmod} we look at evidence for rotational modulation and find that our data are consistent with \ha\ being rotationally modulated. Additionally we discuss the Doppler shifts of the activity indicators and find that both major flares likely originate from a similar location on the stellar surface. Additionally, the red excess seen in the \ha\ wing index is consistently elevated, suggesting ejected material reentering the lower atmosphere. We then compare, in Sect.~\ref{sec:Dis:comparison}, the response of the activity indicators to flares 1 and 2. We present a series of possible scenarios as to why there was such a difference in reaction to the two events. Lastly, we explore the possibility of a CME being associated with the blue asymmetry of flare 2. We find, however, that our data do not support any successful mass ejection of material, but there is some evidence for a failed ejection that later reenters the lower atmosphere causing a disruption to the chromospheric indices.

\subsection{Flaring rates and energies}
\label{sec:Dis:flares}

Given the \textit{TESS} observation period and the observed flare count we calculate a flare rate of 0.818 
flares per day. The sum of their decay times was 5,221\,s for a duty cycle (time flaring 
divided by non-flaring time) of 0.23\,\%. 
Figure~\ref{fig:Tlog} shows that flare~1 and 2, which we covered with multiband,
ground-based photometry, have similar energies and peak luminosities as 
the \textit{TESS} flares observed the month prior. There may be a small separation between high energy, long-duration flares, of which flare 2 is a member, and lower energy flares, of which flare 1 is a 
member. Whether this gap corresponds to some physical property or mechanism remains speculative.

\citealt{Vida2019b}, using \textit{TESS}, measure a flare rate of 1.49 per day on Proxima Centauri with a 
duty cycle of 7.2\,\%. The average energy output of the 72 events observed on Proxima Centauri is 
$11.5\times 10^{30}$\,erg while the average for our observations is $17.8\times 10^{31}$\,erg, that is,
about an order of magnitude more, which is in line with a higher cadence and duty cycle for the
events identified on Proxima Centauri. In a larger 
study of flares observed by \textit{TESS}, \citealt{Gunther2020} find that for M dwarfs, with rotation 
periods $<$ 0.3 d, the flare frequency was between 0.1 and 0.5 per day. In a similar study with Sloan 
Digital Sky Survey, \citealt{Hilton2010} find that for M0 to M1 dwarfs the flare duty cycle was 0.02\,\% 
but went up to 3\,\% for M dwarfs M7 to M9.
Although direct comparisons remain difficult owing to different sensitivities and flare detection
methodology, our findings are generally consistent with those in comparable stars.

\subsection{Localization of flare~2 region}
\label{sec:Dis:loc}

With the information on the Doppler shifts of the chromospheric lines, we can estimate the latitude and longitude of the flaring region. The (rotational)  $\mbox{RV}(t)$
of a surface element as a function of time, $\Delta t$, and stellar latitude, $\phi$, is given by
\begin{equation}
    \mbox{RV}(t) = v\sin{i}\cos{\phi}\sin\left(2\pi\left(\frac{(t-t_0)}{P}+\frac{3}{4}\right) \right)  \; ,
    \label{Eq.:VT}
\end{equation}
where $v\sin i$ is the projected equatorial rotation speed, $P$ is the stellar rotation period,
and $t_0$ is the instance of minimum RV. For a large inclination, this corresponds well to the
instant of appearance of a feature at the limb.

We fit the expression in Eq.~\ref{Eq.:VT} to the H$\alpha$, \hei, \si, and \ca\ shifts of
observations no. 18 to 22, treating $\phi$ and $t_0$ as free parameters. In this way, we
obtained a latitude of $43\pm 10$\,deg and a value of $3.73\pm 0.12$~h for $t_0$, where the
error is estimated using the jackknife method \citep[e.g.,][]{Efron1981}.
We did not use the two observations during and after flare onset as these are the most likely to be contaminated by radial bulk motions, as indicated in Fig.~\ref{fig:WingIndex}.

Given the onset time of flare~2, we estimate that the flare started $70\pm 5$\,deg from the
center of the disk. At this instance, Eq.~\ref{Eq.:VT} yields a shift of $-26\pm 4$\,km\,s$^{-1}$,
which agrees well with the observed RV shift.

The agreement of these two values indicates that the majority of the blueshift of the narrow components originates from the displacement of the active region from the center of the disk and the resulting rotational RV shift rather than bulk motions of flaring material (Figs.\ref{fig:heatmaps} and \ref{fig:Velocity}). It is likely that the bulk motions are better represented by the broad component featured in Fig.~\ref{fig:DG_Flare_Onset}. The observed data and the corotating aspect are similar to an active region with post-flare arcadal loops on our Sun.

\subsection{Minor flares}
\label{sec:Dis:mflares}

Prior to flare 1, Fig.~\ref{fig:Velocity} shows that the \ha\ line is shifted to the red. These redshifts coincide with a series of small flare-like events (Fig.~\ref{fig:miniflares}).  Concurrently with these small flares, the wing index measurement (Fig.~\ref{fig:WingIndex}) of \ha\ shows an enhancement in the red wing  over that of the blue wing. While there are multiple situations in which chromospheric lines can exhibit asymmetries, these red wing enhancements are frequently associated with coronal rain in which the down-falling material emits \ha\ as it heats up upon reentry of the lower atmosphere \citep{Fuhrmeister2018}.  

Starting at two hours into the observation, this redward shift quickly ascends from +10\kms\ to a peak of +26\kms\ within about one hour. This maximum occurs just after the last of the minor flare-like events. It is immediately followed by the activity minimum spectrum and for the rest of the observing run there are no further small flare events. For the rest of our observations, the \ha\ velocity shift value never again reaches the 10\,km\,s$^{-1}$ value. Additionally the slope of the increasing redshift of the \ha\ asymmetry is consistent with a corotating feature. This implies that these minor flares and increasing red asymmetry may be due to an active region moving over the limb of the star just prior to or concurrent with the activity minimum observation. Unfortunately, without data spanning multiple rotation periods, we could not confirm this. We can, however, conclusively rule out any association of the minor flares with flare 1 and 2. If the minor flares were part of the same active region as flare 1 and 2, then they would have occurred while the active region was on the far side of the star and therefore unobservable. 

When comparing the minor flaring across photometric bands (Fig.~\ref{fig:miniflares}), the minor flares are only discernible at short wavelengths ($B$ and $V$ specifically). In the longer wavelength ranges they become indistinguishable from the background. Therefore, if further work is to be done to disentangle possible rotation modulation of \ha\ from the effect of minor flaring (bulk vertical motion; i.e., coronal rain), it should be done with high cadence, simultaneous spectroscopic and photometric ($B$ and $V$) observations.

\subsection{Rotational modulation and Doppler shifts of activity indicators}
\label{sec:Dis:Rotmod}

In Fig.~\ref{fig:TESS_VH}, the trend of \ha\ in the first half of the observing run is similar, if more exaggerated, to that of the light curve of \textit{TESS}, suggesting that the \ha\ index is modulated by rotation. However owing to the onset of flare 2 in the second half of the data set we could not confirm this. There also exists the possibility that \ha\ index has a periodicity twice that of the rotational period of the star as seen in other active M dwarfs (\citealt{Schoefer2019}).

An alternative scenario for the pre-flare absorption dip in \ha\ is supplied by \citet{Jardine2020}. They argue that rapidly rotating and young ($<$ 800\,Myr) stars (such as GJ~3270) are prone to having slingshot prominences. Slingshot prominences are comprised of trapped, cool, gas and appear as absorption transients in \ha. Slingshot prominences are thought to be most common around zero age main-sequence stars \citep{Jardine2020,Cang2020}. If such a prominence was present prior to flare 2, it was likely disrupted or ejected by that flare because no similarly sized absorption transients are seen for the rest of our observational period. Whether this absorption feature in \ha\ is due to a prominence or  the rotational modulation of \ha\ is unclear. 

For two hours after flare 1 (this time frame includes flare 2) all indicators are closely correlated (Fig.~\ref{fig:Velocity}). This suggests that flare 1 and 2 are related and likely originate from the same active region. Therefore, given the determined onset position of flare 2, flare 1 would have occurred at or just over the limb of the star.

After flare 2, the Doppler shifts are consistent with a corotating feature, as detailed in Sect.~\ref{sec:Dis:loc}. At $\sim$6\,h into the observations, all of the Doppler shifts of the activity indicators have returned to their line cores and from this point forward become redshifted (Fig.~\ref{fig:Velocity}). Simultaneously, the activity indicator Doppler shifts diverge from that expected of a corotating feature, suggesting that the post-flare effects in the active region have subsided or are no longer the dominant feature of activity on the disk. Additionally, all the activity indices increase except for \ha\,, which halts its decay and plateaus for $\sim$45\,min (Fig.~\ref{fig:Index}). The wing index (Fig.~\ref{fig:WingIndex}) shows a larger red enhancement at this time than it did during the period of minor flaring. We find these data are consistent with a period of coronal rain, possibly the result of material partially ejected during the onset of flare 2, returning to the star.

\subsection{Comparison of flare 1 and 2}
\label{sec:Dis:comparison}

While the position of flare 2 can be calculated by its after effects, flare 1 is considerably less intense (Table~\ref{table_EF}, Fig.~\ref{fig:muscatsno}) and has no discernible after effects. We must therefore infer its relation to flare 2.
\begin{table} 
    \caption{Enhancement factor of flare 2 from flare 1 by photometric band.}\label{table_EF} 
    \centering 
    \begin{tabular}{c c c c c c} 
        \hline \hline
         \noalign{\smallskip} 
       $B$ & $V$ & $r$ & $i$ & $z$ & mean \\ 
        \noalign{\smallskip} 
        \hline
         \noalign{\smallskip} 
       3.36&2.6&1.9&1.19&...&2.26\\
        \hline 
    \end{tabular} 
\end{table}

Just prior to flare 1, the \ha\ line had reached its maximum redshift value of $\sim$25\,km\,s$^{-1}$ (Fig.~\ref{fig:Velocity}). We previously surmised that this may be the signature of an active region \textasciitilde30~min from going over the limb of the star. At the onset of flare 1, all Doppler shifts of the activity indicators shift toward the blue. After flare 1, these shifts increase until the maximum blueshift occurs at the onset of flare 2. The Doppler shifts of all the activity indicators are well correlated from flare 1 onset to $\sim$90\,min after flare 2 onset, indicating that the source of this shift is the dominant chromospheric feature on the star. 

The calculated location of the active region that spawned flare 2 at the time of the onset of flare 2 is 70$\pm$5\,deg from disk center. It would have taken this active region $\sim$30$\pm$8\,min to arrive at this location from the limb. The timing uncertainty and longer visibility of higher latitudes inclined toward the observer allow that flare~1 originated from the same active region as flare~2. If this active region were the source of flare 1 then flare 1 would have occurred at or near the limb of the star. That we did not see the full blueshifted value of this active region, at that time, could have been due to either a lack of a strong signal or possibly to residual, contaminating effects of the other active region moving over the far limb. This would have been the same active region from whence the minor flares had occurred earlier in the observation period.

This positional argument is supported by the response of the activity indicators (Fig.~\ref{fig:Index}). We conservatively took the flare 2 onset values for $I/I_{r}$ and divided them by the $V$-band flare 2 over flare 1 enhancement (2.6, Table~\ref{table_EF}) giving us a list of expected activity indicator values for the response to flare 1 if it were proportional to flare 2 (Table~\ref{table_AR}). 
\begin{table} 
    \caption{Expected vs. observed activity indicator response to flare 1}\label{table_AR} 
    \centering 
    \begin{tabular}{lccccc} 
        \hline \hline 
         \noalign{\smallskip} 
        & \ha & \hei &  \sii &\si &\ca \\ 
         \noalign{\smallskip} 
        \hline
         \noalign{\smallskip} 
        Flare 2 obs & 1.35&1.29&1.44&1.36&1.06\\
        Flare 1 exp & 1.13&1.12&1.17&1.14&1.02\\
        Flare 1 obs &1.05&1.08&1.15&1.16&1.02\\
        \hline 
    \end{tabular} 
    \footnotesize{\textbf{Note}: Error on figures: 0.02.}
\end{table}
This allowed us to compare the expected with the observed indicator values of flare 1. We found that \sii, \si\ , and \ca\ responded to flare 1 in proportion to their response to flare 2. \ha\ and \hei\, however, did not. Both of these indicators were somewhat weaker in flare 1 than would be expected. A possible explanation for this is preferential absorption. 

Preferential absorption could arise from an extended light path through the mid-upper atmosphere in which \ha\ and \hei\ form. In these regions the temperatures are too hot for the ground states of the sodium and calcium lines, thereby allowing the \sii, \si\ , and \ca\ lines through unhindered whilst absorbing some of the \ha\ and \hei. This extended path would be expected for a source near the limb of the star. While we cannot say for certain that flare 1 occurred on the limb or that it is associated with flare 2, we find the evidence for this case plausible. In this case the observed differences in activity indicator response between flare 1 and 2 would be due to a viewing angle effect.

\subsection{Ejection of material}
\label{sec:Dis:cme}

In our observations we do see a large blue asymmetry that has a broad, asymmetric component. This indicates that some bulk plasma motion was occurring during our observations. However the velocity of these plasma motions was at most 30\,km\,s$^{-1}$, which is only 5\% of the escape velocity. Additionally if this material originated from the same active region as the narrow component then it should also have a $\sim$25\,\kms\ blueshift due to rotation. We therefore conclude that the detected bulk plasma motions did not result in a CME. However, as we already noted in Section~\ref{sec:results:spectroscopy}, about 90 min after the onset of flare 2 there was a change in the decay trend of all chromospheric activity indicators into an increasing trend (except for \ha\ which halts its decay and enters a plateau for $\sim$45\,min). During this time the red wing enhancement is at its peak, superseding that of the earlier flaring period. This would indicate, with the assumption that this red wing enhancement is due to an increase in coronal rain, that a considerable amount of material is falling through the chromosphere. During this same time period there were no indications in the photometric data of further flaring activity. We find our data consistent for either a failed loop ejection or a failed CME. In this scenario material from this event rises into the upper atmosphere before raining down and releasing the kinetic energy into the chromosphere, thereby triggering the increase in activity that we observe.

\section{Conclusions}

We report a series of flares, including a large flare that was followed by a corotating feature, on the  ultra-fast-rotating M4.5\,V star GJ~3270. 
We analyzed 27 spectra taken with CARMENES on 15 December 2018. 
Simultaneous to these observations, photometric observations out in $B$ and $V$ bands from Sierra Nevada and observations in the $r$, $i$, and $z$ bands by MuSCAT2 from Teide were carried out. 
Just prior to our ground-based observations, \textit{TESS} monitored GJ~3270 for 26 d in a row. 

Early in the CARMENES+SNO+MuSCAT2 observing period, a series of minor flaring events were observed along with associated red asymmetries in the \ha\ line. This is consistent with the interpretation that these flares were inducing coronal rain.  
Just prior to the cessation of minor flaring, these red asymmetries increased to the point that the Doppler shift of the residual flux in the \ha\ had nearly reached the $v \sin{i}$ of the star. 
No further minor flares were detected for the rest of the observation period. 
This is consistent with the interpretation that an active region was rotating off the observable disk. 

A flare that was larger than those seen during the earlier period of minor flaring occurred 45 min later. This flare (flare 1) had an unusual reaction from the chromospheric activity indicators. 
Typically \ha\ is the most sensitive line in flaring situations. In this case, however, the sodium D lines appeared to be the most sensitive followed by \hei. This unusual reaction coupled with the location of the next, larger flare (flare 2) suggests that flare 1 occurred at or just over the limb of the star. This difference in reaction of the activity indicators, coupled with the position of the flare, can be due to a number of different scenarios. While preferential absorption through a light path containing more stellar atmosphere is a likely explanation, more such situations would have to be observed to come to any firm conclusions. 

Flare 2 had energies on the order of 10$^{32}$\,erg\,s$^{-1}$. This is comparable to the events detected during the \textit{TESS} observation period. The most noticeable feature of flare 2 is the strong blue asymmetry in \ha\ that persisted for $\sim$90\,min. At flare onset this asymmetry could be separated into a narrow component and a broad component. The broad component was $\sim$15\,{\AA} wide, asymmetric and blueshifted by 30\,km\,s$^{-1}$ from the line core. This broad component was visually evident only at the flare onset and may have persisted into the next exposure at a minimal level for a total duration of $\sim$30\,min. We associate this broad component as indicative of bulk plasma motion. The narrow component was blueshifted by 25.9\,km\,s$^{-1}$ from the line core. This component is what persisted for the $\sim$90\,min for which the blue asymmetry was observed. In other activity indicators (sodium D lines, \hei, \ca) an emission peak was observed that is blueshifted by $\sim$25\,km\,s$^{-1}$ as well. These features then proceeded, well correlated to one another, to shift toward the line core over the subsequent 90 min, while the amplitude of the shifting component decreased. The rate of this shift is consistent with a corotating surface feature that originates at $\sim$65\,deg\ away from disk center with a latitude of $\sim$40\,deg. To our knowledge this is the first time such a feature has been observed on an M dwarf. The data are consistent with the solar analogy of an active region experiencing arcadal loops. 

Approximately 6 h into the observations and 2 h after flare 2 onset, an increase in chromospheric activity indicators occurred. Associated with this increase was also an elevated period of red asymmetries associated with many of the activity lines. These red asymmetries were larger than those observed earlier during the period of minor flaring. During this same time period, however, no flaring activity was detected in any of the photometric bands. Given that bulk plasma motions were detected during the onset of flare 2, these data are consistent with a period of intense coronal rain, possibly resulting from the reentry of ejected material. 

We conclude that the main phenomenon behind our observations was a corotating feature analogous to an active region with arcadal loops. Beyond that, while we have not conclusively shown that a failed CME occurred or that flare regions have a temperature stratification, we have seen sufficient suggestive evidence to warrant further simultaneous, spectroscopic, and photometric observations of fast-rotating M dwarfs.


\begin{acknowledgements}
This project was funded principally by the Deutsche Forschungsgemeinschaft through the Major Research Instrumentation Programme and Research Unit FOR2544 ``Blue Planets around Red Stars''.
    
 CARMENES is an instrument at the Centro Astron\'omico Hispano-Alem\'an (CAHA) at Calar Alto (Almer\'{\i}a, Spain), operated jointly by the Junta de Andaluc\'ia and the Instituto de Astrof\'isica de Andaluc\'ia (CSIC).
  
  [The authors wish to express their sincere thanks to all members of the Calar Alto staff for their expert support of the instrument and telescope operation.]
  
  CARMENES was funded by the Max-Planck-Gesellschaft (MPG), 
  the Consejo Superior de Investigaciones Cient\'{\i}ficas (CSIC),
  the Ministerio de Econom\'ia y Competitividad (MINECO) and the European Regional Development Fund (ERDF) through projects FICTS-2011-02, ICTS-2017-07-CAHA-4, and CAHA16-CE-3978, 
  and the members of the CARMENES Consortium 
  (Max-Planck-Institut f\"ur Astronomie,
  Instituto de Astrof\'{\i}sica de Andaluc\'{\i}a,
  Landessternwarte K\"onigstuhl,
  Institut de Ci\`encies de l'Espai,
  Institut f\"ur Astrophysik G\"ottingen,
  Universidad Complutense de Madrid,
  Th\"uringer Landessternwarte Tautenburg,
  Instituto de Astrof\'{\i}sica de Canarias,
  Hamburger Sternwarte,
  Centro de Astrobiolog\'{\i}a and
  Centro Astron\'omico Hispano-Alem\'an), 
  with additional contributions by the MINECO, 
  the Klaus Tschira Stiftung, 
  the states of Baden-W\"urttemberg and Niedersachsen
  and by the Junta de Andaluc\'{\i}a.
  
  We acknowledge financial support from the Agencia Estatal de Investigaci\'on of the Ministerio de Ciencia, Innovaci\'on y Universidades and the ERDF through projects 
  PID2019-109522GB-C5[1:4]/AEI/10.13039/501100011033    
  and PGC2018-098153-B-C33                                      

and the Centre of Excellence ``Severo Ochoa'' and ``Mar\'ia de Maeztu'' awards to the Instituto de Astrof\'isica de Canarias (SEV-2015-0548), Instituto de Astrof\'isica de Andaluc\'ia (SEV-2017-0709), and Centro de Astrobiolog\'ia (MDM-2017-0737), the Generalitat de Catalunya/CERCA programme,
JSPS KAKENHI via grants JP18H01265 and JP18H05439, and JST PRESTO via grant JPMJPR1775.

This work was based on data from the CARMENES data archive at CAB (CSIC-INTA).
  
Data were partly collected
with the 150\,cm and 90\,cm telescopes at the Observatorio de Sierra Nevada (SNO) operated by the Instituto de Astrof\'\i fica de Andaluc\'\i a (IAA-CSIC).

This article is based on observations made with the MuSCAT2 instrument, developed by ABC, at Telescopio Carlos S\'nchez operated on the island of Tenerife by the IAC in the Spanish Observatorio del Teide.

\end{acknowledgements}


\bibliographystyle{aa}

\appendix

\section{Tables}

\begin{table*} 
    \caption{CARMENES observations log.}\label{TB:timetable} 
    \centering 
    \begin{tabular}{cccc} 
        \hline \hline 
      Obs\# & BJD-BJD$_0$ [h] & BJD [d] \\
      \hline
      1 & 0.0 & 2458468.3156583 \\  
      2 & 0.27 & 2458468.3269369  \\  
      3 & 0.55 & 2458468.3384585  \\  
      4 & 0.83 & 2458468.3503736 \\  
      5 & 1.11 & 2458468.3618431  \\  
      6 & 1.37 & 2458468.3727224  \\  
      7 & 1.65 & 2458468.3846028  \\  
      8 & 1.91 & 2458468.3954126  \\  
      9 & 2.18 & 2458468.4066391  \\  
      10 & 2.45 & 2458468.4177151 \\  
      11 & 2.73 & 2458468.4293293  \\  
      12 & 2.99 & 2458468.4402606   \\  
      13 & 3.29 & 2458468.4525866  \\  
      14 & 3.55 & 2458468.4636279  \\  
      15 & 3.83 & 2458468.4752884  \\  
      16 & 4.09 & 2458468.4861445  \\  
      17 & 4.37 & 2458468.4977761  \\  
      18 & 4.63 & 2458468.5084354  \\  
      19 & 5.08 & 2458468.5271732  \\  
      20 & 5.34 & 2458468.5380409  \\  
      21 & 5.6 & 2458468.5491227  \\  
      22 & 5.88 & 2458468.5606096  \\  
      23 & 6.16 & 2458468.57238  \\  
      24 & 6.43 & 2458468.5836759  \\  
      25 & 6.71 & 2458468.5950413  \\  
      26 & 6.98 & 2458468.6063893  \\  
      27 & 7.26 & 2458468.6182928  \\  
      28 & 7.55 & 2458468.6300401 \\  
        \hline 
    \end{tabular} 
\end{table*}

\begin{table*} 
    \caption{Activity indicator index values.}
    \label{indextable} 
    \centering 
    \begin{tabular}{cccccccccccc} 
        \hline \hline 
       Obs\# & BJD-BJD$_0$ [h] & \ha  &  $\delta$\ha & \hei &  $\delta$\hei &  \sii &  $\delta$\sii &  \si &  $\delta$\si &  \ca&  $\delta$\ca \\
       \hline
       1 & 0.00 & 2.7 & 0.01 & 1.14 & 0.02 & 0.51 & 0.02 & 0.53 & 0.02 & 0.96 & 0.003 \\  
       2 & 0.27 & 2.67 & 0.01 & 1.19 & 0.02 & 0.51 & 0.02 & 0.52 & 0.02 & 0.96 & 0.003 \\  
       3 & 0.55 & 2.79 & 0.01 & 1.19 & 0.02 & 0.53 & 0.01 & 0.55 & 0.02 & 0.96 & 0.002 \\  
       4 & 0.83 & 2.90 & 0.01 & 1.28 & 0.02 & 0.53 & 0.01 & 0.53 & 0.02 & 0.97 & 0.003 \\  
       5 & 1.11 & 2.75 & 0.01 & 1.22 & 0.02 & 0.54 & 0.01 & 0.55 & 0.01 & 0.96 & 0.002 \\  
       6 & 1.37 & 2.82 & 0.01 & 1.26 & 0.02 & 0.55 & 0.01 & 0.55 & 0.01 & 0.97 & 0.002 \\  
       7 & 1.65 & 2.75 & 0.01 & 1.24 & 0.02 & 0.53 & 0.01 & 0.52 & 0.01 & 0.97 & 0.002 \\  
       8 & 1.91 & 2.67 & 0.01 & 1.23 & 0.02 & 0.51 & 0.01 & 0.54 & 0.01 & 0.96 & 0.002 \\  
       9 & 2.18 & 2.64 & 0.01 & 1.18 & 0.02 & 0.57 & 0.01 & 0.56 & 0.01 & 0.96 & 0.002 \\  
       10 & 2.45 & 2.60 & 0.01 & 1.18 & 0.02 & 0.53 & 0.01 & 0.54 & 0.01 & 0.96 & 0.002 \\  
       11 & 2.73 & 2.55 & 0.01 & 1.20 & 0.02 & 0.53 & 0.01 & 0.55 & 0.01 & 0.96 & 0.002 \\  
       12 & 2.99 & 2.56 & 0.01 & 1.20 & 0.01 & 0.50 & 0.01 & 0.54 & 0.01 & 0.96 & 0.002 \\  
       13 & 3.29 & 2.41 & 0.01 & 1.15 & 0.01 & 0.49 & 0.01 & 0.5 & 0.01 & 0.95 & 0.002 \\  
       14 & 3.55 & 2.53 & 0.01 & 1.25 & 0.01 & 0.57 & 0.01 & 0.58 & 0.01 & 0.96 & 0.002 \\  
       15 & 3.83 & 2.56 & 0.01 & 1.23 & 0.01 & 0.54 & 0.01 & 0.55 & 0.01 & 0.96 & 0.002 \\  
       16 & 4.09 & 2.55 & 0.01 & 1.20 & 0.02 & 0.51 & 0.01 & 0.52 & 0.01 & 0.95 & 0.002 \\  
       17 & 4.37 & 3.26 & 0.01 & 1.49 & 0.02 & 0.71 & 0.01 & 0.69 & 0.01 & 1.01 & 0.002 \\  
       18 & 4.63 & 3.50 & 0.01 & 1.41 & 0.02 & 0.70 & 0.01 & 0.67 & 0.01 & 1.04 & 0.003 \\  
       19 & 5.08 & 3.69 & 0.01 & 1.43 & 0.02 & 0.72 & 0.01 & 0.67 & 0.01 & 1.04 & 0.003 \\  
       20 & 5.34 & 3.54 & 0.01 & 1.34 & 0.02 & 0.61 & 0.01 & 0.60 & 0.02 & 1.01 & 0.003 \\  
       21 & 5.60 & 3.27 & 0.01 & 1.23 & 0.02 & 0.63 & 0.02 & 0.62 & 0.02 & 1.0 & 0.003 \\  
       22 & 5.88 & 3.12 & 0.02 & 1.21 & 0.03 & 0.57 & 0.02 & 0.56 & 0.02 & 0.98 & 0.003 \\  
       23 & 6.16 & 3.06 & 0.01 & 1.24 & 0.02 & 0.6 & 0.02 & 0.56 & 0.02 & 0.98 & 0.003 \\  
       24 & 6.43 & 3.06 & 0.01 & 1.28 & 0.02 & 0.63 & 0.02 & 0.60 & 0.02 & 0.98 & 0.002 \\  
       25 & 6.71 & 3.05 & 0.01 & 1.25 & 0.03 & 0.56 & 0.02 & 0.55 & 0.02 & 0.98 & 0.003 \\  
       26 & 6.98 & 2.96 & 0.02 & 1.19 & 0.03 & 0.62 & 0.02 & 0.59 & 0.02 & 0.97 & 0.003 \\  
       27 & 7.26 & 2.93 & 0.02 & 1.22 & 0.04 & 0.65 & 0.03 & 0.61 & 0.03 & 0.97 & 0.004 \\  
       28 & 7.55 & 2.85 & 0.02 & 1.19 & 0.04 & 0.66 & 0.03 & 0.57 & 0.03 & 0.96 & 0.004 \\  
        \hline 
    \end{tabular} 
    \footnotesize{BJD$_0$ = 2458468.3156583[BJD]}
\end{table*}

\begin{table*} 
    \caption{\ha\ wing index values.}\label{wingindextable} 
    \centering 
    \begin{tabular}{ccccccccc} 
        \hline \hline 
       Obs\# & BWI  &  $\delta$BWI & RWI &  $\delta$RWI &  BWI-e &  $\delta$BWI-e &  RWI-e &  $\delta$RWI-e \\ 
       \hline
       1 & 0.971 & 0.008 & 1.022 & 0.007 & 0.991 & 0.007 & 1.003 & 0.006 \\  
       2 & 0.963 & 0.007 & 1.006 & 0.006 & 0.981 & 0.006 & 0.984 & 0.006 \\  
       3 & 0.971 & 0.006 & 1.055 & 0.006 & 0.985 & 0.006 & 1.002 & 0.005 \\  
       4 & 1.009 & 0.006 & 1.067 & 0.006 & 1.000 & 0.006 & 1.003 & 0.005 \\  
       5 & 0.969 & 0.006 & 1.026 & 0.006 & 0.988 & 0.006 & 0.987 & 0.005 \\  
       6 & 1.002 & 0.006 & 1.037 & 0.005 & 0.998 & 0.006 & 1.005 & 0.005 \\  
       7 & 0.995 & 0.006 & 1.005 & 0.005 & 0.999 & 0.006 & 0.986 & 0.005 \\  
       8 & 0.988 & 0.005 & 0.99 & 0.005 & 0.991 & 0.005 & 0.987 & 0.005 \\  
       9 & 1.001 & 0.006 & 0.993 & 0.005 & 1.004 & 0.006 & 0.986 & 0.005 \\  
       10 & 0.982 & 0.006 & 0.994 & 0.005 & 0.993 & 0.005 & 0.988 & 0.005 \\  
       11 & 0.989 & 0.005 & 0.998 & 0.005 & 0.995 & 0.005 & 0.997 & 0.004 \\  
       12 & 1.019 & 0.005 & 1.048 & 0.005 & 1.014 & 0.005 & 1.020 & 0.004 \\  
       13 & 1.000 & 0.005 & 1.000 & 0.005 & 1.000 & 0.005 & 1.000 & 0.004 \\  
       14 & 1.014 & 0.005 & 1.049 & 0.005 & 1.000 & 0.005 & 1.019 & 0.004 \\  
       15 & 1.018 & 0.005 & 1.021 & 0.004 & 1.007 & 0.005 & 1.001 & 0.004 \\  
       16 & 0.983 & 0.005 & 0.981 & 0.004 & 0.981 & 0.005 & 0.979 & 0.004 \\  
       17 & 1.241 & 0.006 & 1.171 & 0.005 & 1.109 & 0.005 & 1.082 & 0.005 \\  
       18 & 1.122 & 0.006 & 1.073 & 0.006 & 1.04 & 0.006 & 1.023 & 0.005 \\  
       19 & 1.049 & 0.006 & 1.08 & 0.006 & 1.008 & 0.006 & 1.012 & 0.005 \\  
       20 & 1.018 & 0.007 & 1.103 & 0.006 & 1.000 & 0.006 & 1.011 & 0.006 \\  
       21 & 0.994 & 0.007 & 1.099 & 0.007 & 0.989 & 0.007 & 1.022 & 0.006 \\  
       22 & 0.979 & 0.008 & 1.029 & 0.007 & 0.979 & 0.007 & 0.995 & 0.006 \\  
       23 & 0.999 & 0.007 & 1.047 & 0.006 & 1.000 & 0.006 & 1.028 & 0.006 \\  
       24 & 0.999 & 0.006 & 1.036 & 0.006 & 1.000 & 0.006 & 1.011 & 0.005 \\  
       25 & 1.002 & 0.007 & 1.014 & 0.007 & 0.993 & 0.007 & 1.004 & 0.006 \\  
       26 & 0.985 & 0.008 & 1.002 & 0.007 & 0.984 & 0.008 & 0.998 & 0.007 \\  
       27 & 0.994 & 0.010 & 1.031 & 0.009 & 0.998 & 0.009 & 1.015 & 0.008 \\  
       28 & 0.973 & 0.010 & 1.012 & 0.009 & 0.990 & 0.010 & 1.000 & 0.009 \\  
        \hline 
    \end{tabular} 
\end{table*}

\end{document}